\documentclass[12pt,preprint]{aastex}
\usepackage{epsf}

\newcommand{\be}{\begin{displaymath}}
\newcommand{\ee}{\end{displaymath}}
\newcommand{\bea}{\begin{eqnarray}}
\newcommand{\eea}{\end{eqnarray}}

\shortauthors{Pavel Denissenkov}
\shorttitle{A Model of Magnetic Braking of Solar Rotation}

\begin{document}

\title{A MODEL OF MAGNETIC BRAKING OF SOLAR ROTATION
       THAT SATISFIES OBSERVATIONAL CONSTRAINTS}

\author{Pavel A. Denissenkov}
\affil{Department of Physics \& Astronomy, University of Victoria,
       P.O.~Box 3055, Victoria, B.C., V8W~3P6, Canada}
       \email{pavel.denisenkov@gmail.com}
 
\begin{abstract}
The model of magnetic braking of solar rotation considered by \cite{chmg93} has been modified so that
it is able to reproduce for the first time the rotational evolution of both the fastest and slowest rotators among solar-type stars
in open clusters of different ages, without coming into conflict with other observational constraints, such as
the time evolution of the atmospheric Li abundance in solar twins and the thinness of the solar tachocline.
This new model assumes that rotation-driven turbulent diffusion, which is thought to amplify the viscosity and
magnetic diffusivity in stellar radiative zones, is strongly anisotropic with the horizontal components of the transport coefficients strongly
dominating over those in the vertical direction. Also taken into account is the poloidal field decay
that helps to confine the width of the tachocline at the solar age. The model's properties are investigated
by numerically solving the azimuthal components of the coupled momentum and magnetic induction equations
in two dimensions using a finite element method.
\end{abstract} 

 
\section{Introduction}
\label{sec:intro}

Helioseismology has revealed that the Sun's radiative core rotates nearly uniformly, at least down to the radius $r\sim 0.2\,R_\odot$,
with a rate that is close to the mean rotation rate of its convective envelope (e.g., \citealt{tst95,cea03}).
It is also known that the Sun, like other solar-type stars, has been experiencing an effective braking of its
surface rotation via a magnetized solar (stellar) wind (\citealt{s62,wd67,k88}). Because the magnetic field lines that sling charged particles 
into space are rooted in the photosphere, it is the envelope rotation that should be decelerated by the wind torque,
while the conservation of angular momentum in the radiative core should keep it rapidly rotating. This expectation appears to be
in a stark contrast with the measurements of internal solar rotation, which indicate that the difference in the angular velocity
between the Sun's radiative core and convective envelope must have been erased by some angular momentum redistribution mechanism.        
A similar conclusion can be reached from comparisons of the rotation period distributions for solar-type stars in open clusters of
different ages. It turns out that the core and envelope rotation have to be coupled with a characteristic timescale of
angular momentum transfer between them increasing from a few Myr to nearly a hundred Myr between the fastest and slowest
rotators (\citealt{dea09}; cf. \citealt{iea07}). A simple model of rotational evolution of solar-type stars with angular momentum transport
described as a purely diffusive process shows that this coupling time interval roughly corresponds to a diffusion coefficient
decreasing from $10^6$\,cm$^2$\,s$^{-1}$ to $5\times 10^4$\,cm$^2$\,s$^{-1}$. However, such large diffusion
coefficients would lead to rates of surface Li destruction strongly exceeding the trend observed in the Sun and its twins (Section \ref{sec:difmod}).
Therefore, we have to admit that the angular momentum redistribution in solar-type stars is unlikely to be accompanied by
an equally rapid mass transfer. This restricts our search for possible mechanisms of angular momentum transport
in solar-type stars to those employing waves and magnetic fields.

Presently, the most popular mechanisms of angular momentum redistribution in the Sun's radiative core are 
the smoothing of its rotation profile by internal gravity waves (g-modes) generated by turbulent eddies in the convective envelope (\citealt{cht05})
and magnetic braking of differential rotation (\citealt{chmg92,chmg93,s99,s02}). Using a simplified analytical
prescription for the spectrum of internal gravity waves and other approximations, \cite{cht05}
have succeeded in obtaining a nearly flat rotation profile in the present-day Sun simultaneously with a time evolution of
its internal rotation consistent with the observed solar Li abundance. However, because of the complexity of physical
processes involved in the generation, propagation and dissipation of internal gravity waves in the Sun, it is difficult
to assess the validity of this solution. For instance, it would not be valid if a flatter spectrum,
like that computed by \cite{rg05}, were used. Non-linear wave-wave interactions, as well as the interaction of gravity
waves with a magnetic field, could also change the solution drastically (\citealt{rmgg08,rmg09}). Even in a case
very similar to that considered by \cite{cht05} the question arises as to why gravity waves do not disturb the uniform
solar rotation? Indeed, the power of gravity waves generated in the present-day Sun should approximately be the same as it was
in the young Sun; consequently, given their peculiar anti-diffusive nature in shear flows (e.g., \citealt{p77,r98}), gravity waves should have
forced the solar rotation profile away from the uniform one (\citealt{dea08}). However, this has not occurred.

For a magnetic field configuration to influence differential rotation of the solar radiative core, it must have a non-vanishing azimuthal (toroidal) component.
To achieve this, it is usually assumed that there is an axially symmetric poloidal magnetic field frozen into the plasma inside the radiative core
which can be stretched around the rotation axis by the differential rotation itself to form the necessary toroidal field.
As far as the origin of the poloidal field is concerned, it is widely believed that the field was either inherited from a protostellar cloud or generated
by a convective dynamo during the Sun's pre-MS evolution when its convective envelope occupied a much larger volume.
\cite{s99} proposed the original alternative explanation that the poloidal field might be constantly replenished by radially stretching the toroidal field in
a dynamo-like cyclic process, the necessary radial plasma displacements being caused by the non-axisymmetric ($m=1$) kink
instability of the toroidal field (the Tayler-Spruit dynamo). 

Magnetic braking is implemented by the azimuthal component of the magnetic tension term
of the Lorentz force. Given that its corresponding engagement time is of order several Alfv\'{e}n timescales for the poloidal field,
magnetic breaking appears to be very efficient even for a small field amplitude (\citealt{mw87}).
The only problem associated with this mechanism is that it leaves behind a torus-shaped
``dead'' zone inside the radiative core with the angular velocity increasing toward the center of the torus cross-section.
This is a direct consequence of Ferraro's law of isorotation (rotation that keeps the angular velocity constant along each field line) (\citealt{f37})
and the presence of a constant dipole poloidal field. To solve this problem, \cite{chmg93} (hereafter, referred to as CHMG93)
and other researchers (e.g., \citealt{rk96}) artificially amplified the viscous transport with the justification that rotation and associated
hydrodynamic instabilities should lead to turbulence, and hence that the microscopic viscosity should be augmented by
a much stronger turbulent viscosity. However, the minimum value of the enhanced viscosity ($5\times 10^4$\,cm$^2$\,s$^{-1}$)
with which it is possible to get a nearly flat rotation profile in the present-day Sun results in too much Li being transported
below the bottom of convective envelope and thus destroyed, inconsistent with the observed
solar Li abundance. Therefore, the proposed isotropic viscosity enhancement cannot be considered a satisfactory solution.

Only recently has the physics-based modeling of angular momentum transport in solar-type stars
begun to use the extensive data sets of rotation periods for solar analogs in open clusters as an observational constraint.
The main goal of previous studies was to demonstrate that a particular model could (or could not) produce the solid-body rotation of the Sun.
In particular, \cite{eea05} showed that the Tayler-Spruit dynamo could account for the flat rotation profile of the Sun.
However, the correct model must also explain the cause of the transition from the solid-body rotational evolution of the fastest rotators 
to the evolution wherein a radial differential rotation is sustained for a hundred Myr in the slowest rotators.
It turns out that the Tayler-Spruit dynamo cannot pass this observational test because it always enforces a nearly uniform rotation
in solar-type stars, no matter how slowly they rotate (\citealt{dea09}).

Another important problem that the correct mechanism of angular momentum transport in the Sun has to explain is the thinness of
the solar tachocline. The tachocline is a layer in which the differential rotation (both radial and latitudinal) of the convective envelope
sharply changes to the nearly solid-body rotation of the radiative core (\citealt{sz92}). Its thickness, as measured by helioseismology,
is only about 4\% of the solar radius (\citealt{ba03}). It has been suggested that no purely hydrodynamic mechanism can explain
its existence and that a large-scale magnetic field in the radiative core is therefore needed to confine the tachocline structure
(\citealt{gmi98,rk07}).

To summarize: because large-scale magnetic fields have certainly been playing crucial roles during different phases of the Sun's rotational evolution
--- such as the synchronizing of rotation of the young Sun and its surrounding protoplanetary disk (the disk locking) (e.g., \citealt{shea94,mp05}), 
providing the leverage for the solar wind,
confining the tachocline, being generated through a convective dynamo and then driving the solar activity --- we believe that it is worthwhile
to develop a mechanism of magnetic breaking of differential rotation of the Sun's radiative core.

In this paper, we elaborate upon and extend the magnetic braking mechanism that was originally proposed by \cite{mw87} and then studied
in detail numerically by CHMG93 followed by \cite{rk96}. Our only modification of the mechanism is to assume that the turbulence
induced by rotation-driven hydrodynamic instabilities in the Sun's radiative core is highly anisotropic with its corresponding
horizontal component of turbulent viscosity $\nu_{\rm h}$ strongly dominating over  that in the vertical direction $\nu_{\rm v}$.
The hypothesis of anisotropic turbulent diffusion in stellar radiative zones was first advanced by \cite{z92} and it had since been productively
used by many other researchers to study rotational mixing in both MS and red giant branch stars (\citealt{tz97,tea97,m03,pea03,pea06}).
Initially, it was even considered to explain the thinness of the solar tachocline by \cite{sz92}. We will show that
this hypothesis alone permits a solution of almost all of the aforementioned problems associated with this particular mechanism of
magnetic braking, such as the Li problem (a sufficiently small value of the diffusion coefficient $\nu_{\rm v}$ can be chosen so that
it does not lead to excessive Li destruction), the dead-zone problem, and the difference in spin-down of
the fastest and slowest rotators among solar-type stars in open clusters.

The paper is organized as follows. In Section \ref{sec:simple}, we briefly discuss simple models of rotational evolution of
solar-type stars and review the main results that were obtained by comparison of the model predictions with the observed period distributions for
solar counterparts in open clusters. In Section \ref{sec:chmg93}, we reproduce the 2D MHD computations of CHMG93 and show
that they cannot account for the spin-down of the fastest rotators in open clusters, especially when one reduces the vertical 
viscosity to a value more or less compatible with the surface Li abundance evolution in solar twins. In Section \ref{sec:anisotrop},
we present our new dynamic model of magnetic braking with the anisotropic turbulent diffusion ($\nu_{\rm h}\gg\nu_{\rm v}$) and discuss its advantages compared
to the original model. Section \ref{sec:tacho} describes a number of stationary solutions that address the problem of the penetration of 
the latitudinal differential rotation from the bottom of convective envelope into the radiative interior in the present-day Sun.
Finally, we summarize our conclusions in Section \ref{sec:concl}.

\section{Simple Models of Rotational Evolution of Solar-Type Stars}
\label{sec:simple}

The simple models of rotational evolution of solar-type stars --- i.e., the double-zone model and the purely diffusive one 
--- have been discussed in detail and compared with each other by \cite{dea09}. Here, we will only briefly describe the models and summarize
the main results obtained by comparing their predictions with distributions of angular velocity of the (assumed) rigidly rotating convective envelope for
solar analogs in open clusters, $\Omega_{\rm e} = 2\pi/P$, which are derived from the corresponding observed rotation period ($P$) distributions. 
We will also show that the purely diffusive model fails to simultaneously comply with the observational constraints imposed by
the spin-down of solar analogs in open clusters and the time evolution of the surface Li abundance in solar twins.
Although the simple models do not identify the physics behind the internal transport of
angular momentum in solar-type stars, they nevertheless provide us with useful estimates of characteristic timescales of relevant
processes and their dependence on the rotation rate. Therefore, they can serve as a good starting point for our analysis.

\subsection{The Pre-Main Sequence Disk Locking}

In Fig.~\ref{fig:f1}, crosses with vertical error bars represent the upper 90th and lower 10th percentiles of
the $\Omega_{\rm e}$ distributions for stars having masses in the interval $0.9\la M/M_\odot\la 1.1$ sampled from open clusters with ages
between 2 Myr and 600 Myr (for details, see \citealt{dea09}). 
The data for the youngest clusters set up the initial $\Omega_{\rm e}$ distributions for rotational evolution
computations. Of course, these are not the true initial distributions inherited by the stars during their formation epoch because they
have already been modified (reduced) by a disk-locking process compared to what they would have been if the stars' pre-MS evolution had taken place
in complete isolation. The fact is that after its birth a protostar continues to contract before it settles down on the zero-age MS (ZAMS)
at an age of about 40 Myr (for $M\approx 1\,M_\odot$). During this period of time, the star's radius $R$ and total moment of inertia $I$ are decreasing, hence
its $\Omega_{\rm e}$ should be growing as a consequence of angular momentum conservation. Contrary to this, it is the surface angular velocity
rather than the angular momentum that is observed to be nearly preserved in contracting pre-MS stars (e.g., \citealt{rws04}). 
This is usually explained by a complex magnetic interaction that pumps out
angular momentum from the protostar to its surrounding protoplanetary dust disk (e.g., \citealt{shea94,mp05}). As a result, the magnetic disk locking
keeps $\Omega_{\rm e}$ nearly constant (this is modeled by horizontal fragments of dotted curves in Fig.~\ref{fig:f1}) for a period of time up to 10\,--\,20 Myr
at most before the disk disappears. After that, the star's residual contraction and conservation of angular momentum cause an increase of its $\Omega_{\rm e}$
that continues until the star reaches the ZAMS.

\subsection{The Surface Loss of Angular Momentum}

\begin{figure}
\epsfxsize=13cm
\epsffile [40 180 480 595] {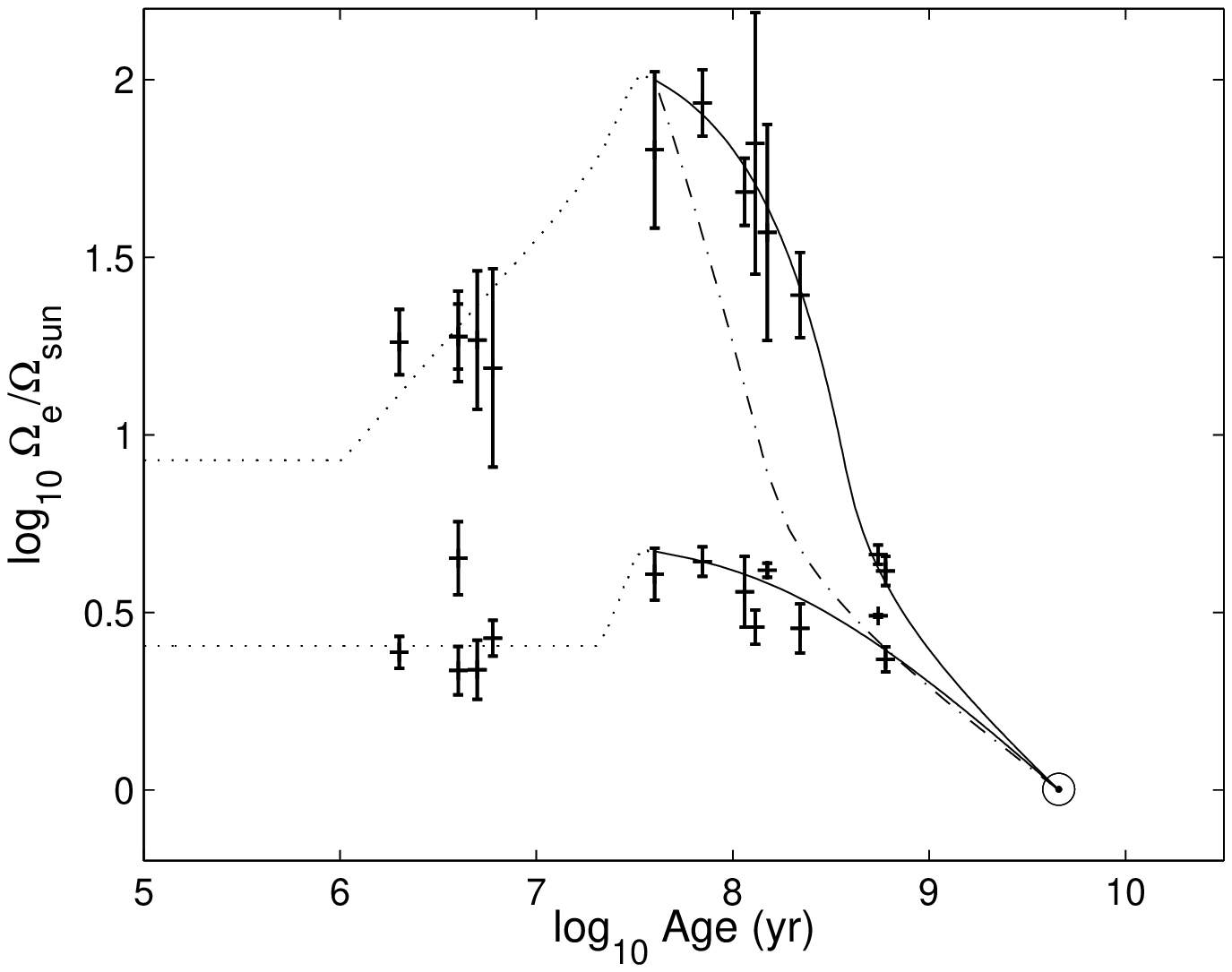}
\caption{Spin-down of the Sun's convective envelope by the torque from a magnetized wind (equation \ref{eq:jdot})
         for the two limiting cases in which the envelope rotation is either totally decoupled or, alternatively, rigidly coupled to that of
         Sun's radiative core (solid curves) and for the case in which the moment of inertia of the uniformly rotating
         envelope, now consisting of the convective zone plus a fraction of its adjacent radiative zone, increases with time 
         according to the law (\ref{eq:plaw}) (dot-dashed curve). Dotted curves (partially overlapped by the solid curves) represent the Sun's
         solid-body rotational evolution computed using the double-zone model with the coupling time $\tau_{\rm c} = 1$ Myr.
         Their initial horizontal fragments correspond to the pre-MS disk-locking phase.
         In this and the following plots, the modeled evolution of the angular velocity of the Sun's convective envelope
         is compared with the upper 90th and lower 10th percentiles of the $\Omega_{\rm e}$ distributions for solar analogs
         in open clusters of different ages (crosses with vertical errorbars) taken from \cite{dea09}.
         }
\label{fig:f1}
\end{figure}

From the moment when the young Sun arrives at the ZAMS ($t_{\rm ZAMS}\approx 40$ Myr) until the Sun's present-day age of $t_\odot = 4.57$ Gyr,
its internal structure does not change significantly. In particular, the dimensionless moment of inertia of the Sun's convective envelope 
$i_{\rm e} \equiv I_{\rm e}/(R_\odot^2 M_\odot)\approx 0.0105$ remains constant to within 6\%. 
Note that $i_{\rm e}$  amounts only to 14\% of the Sun's total moment of
inertia $i\equiv I/(R_\odot^2M_\odot)\approx 0.074$. If there were no angular momentum transfer from the Sun's radiative core 
to its convective envelope then the only process involved
in the Sun's rotational evolution during its life on the MS ($t_{\rm ZAMS}\leq t\leq t_\odot$)
would be the braking of envelope rotation by the magnetized solar wind.
The corresponding rate of angular momentum loss is often approximated by the following equation:
\bea
\dot{J} = \frac{d}{dt}\left(I_{\rm e}\Omega_{\rm e}\right)\approx I_{\rm e}\frac{\Omega_{\rm e}}{dt}\approx -K_{\rm w}\sqrt{\frac{R/R_\odot}{M/M_\odot}}
\,\min\left(\Omega_{\rm e}\Omega_{\rm sat}^2,\,\Omega_{\rm e}^3\right),
\label{eq:jdot}
\eea
where $K_{\rm w}$ is a constant calibrated to give $\Omega_{\rm e}(t_\odot) = \Omega_\odot = 2.86\times 10^{-6}$ rad\,s$^{-1}$ (for $P_\odot = 25.4$ days),
and $\Omega_{\rm sat}$ is a mass-dependent velocity above which the wind gets saturated (\citealt{chea95,kea97,aps03}). 
The value of $\Omega_{\rm sat} = 8\,\Omega_\odot$ has been adjusted
so that the upper solid curve in Fig.~\ref{fig:f1} simulating the solid-body rotational evolution of
rapidly rotating open-cluster solar-type stars could fit the upper 90th percentiles of
their $\Omega_{\rm e}$ distributions as closely as possible (\citealt{dea09}).

The differential equation (\ref{eq:jdot}) can easily be solved. 
If the initial angular velocity $\Omega_{\rm e,ZAMS}\equiv\Omega_{\rm e}(t_{\rm ZAMS})$ 
exceeds the saturation threshold $\Omega_{\rm sat}$ then the solution consists of two different parts.
The first one, which is valid for $t_{\rm ZAMS}\leq t\leq t_{\rm sat}$, is
\bea
\Omega_{\rm e} = \Omega_{\rm e,ZAMS}\exp\left[-(t-t_{\rm ZAMS})/\tau_{\rm w}\right],
\label{eq:exp}
\eea
where the wind braking timescale is $\tau_{\rm w} = (t_\odot - t_{\rm ZAMS})/(A/2+B)$, while
$t_{\rm sat} = t_{\rm ZAMS} + B\tau_{\rm w}$ is the time when $\Omega_{\rm e}$ has decreased to a value of $\Omega_{\rm sat}$.
To shorten the equations, we have used the notations $A\equiv (\Omega_{\rm sat}/\Omega_\odot)^2-1$ and
$B\equiv\ln\,(\Omega_{\rm e,ZAMS}/\Omega_{\rm sat})$.
The exponential decay of $\Omega_{\rm e}$ is replaced by the Skumanich relation (\citealt{s72})
\bea 
\Omega_{\rm e} = \frac{\Omega_\odot}{\sqrt{1 + 2\,[(A/2+B)/(1+A)]\,(t-t_\odot)/(t_\odot-t_{\rm ZAMS})}}
\label{eq:skuma}
\eea
for the time interval $t_{\rm sat} < t \leq t_\odot$.
Finally, the solar-calibrated wind constant has to be equal to
\bea
K_{\rm w} = \frac{R_\odot^2M_\odot}{\Omega_\odot^2}\,\frac{(A/2+B)}{(1+A)}\,\frac{i_{\rm e}}{t_\odot-t_{\rm ZAMS}}.
\label{eq:kw}
\eea
Alternatively, if the star arrives at the ZAMS with $\Omega_{\rm e,ZAMS} \leq \Omega_{\rm sat}$ then its surface angular velocity
will be declining according to the Skumanich law from the very beginning. The above equations
can still be used provided that one substitutes $\Omega_{\rm sat} = \Omega_{\rm e,ZAMS}$ in all of them.

Equations (\ref{eq:exp})\,--\,(\ref{eq:skuma}) suffice to describe the Sun's spin-down for the two limiting cases
in which its core and envelope rotation are either completely decoupled or, alternatively, rigidly coupled. For the first case, 
which is only of academic interest because it is in conflict with
uniform rotation in the solar interior, equation (\ref{eq:kw}) gives $K_{\rm w} = 4.60\times 10^{46}$\,cm$^2$\,g\,s
for $\Omega_{\rm e,ZAMS} = 100\,\Omega_\odot$, and $K_{\rm w} = 4.13\times 10^{46}$\,cm$^2$\,g\,s 
for  $\Omega_{\rm e,ZAMS} = 4.7\,\Omega_\odot$
(the upper and lower solid curves in Fig.~\ref{fig:f1}). In the second case, which appears to adequately describe
the fastest rotators, we have to substitute the total
moment of inertia $i$ instead of $i_{\rm e}$ into (\ref{eq:kw}), resulting in
$K_{\rm w} = 3.24\times 10^{47}$\,cm$^2$\,g\,s and $K_{\rm w} = 2.91\times 10^{47}$\,cm$^2$\,g\,s for the same initial
angular velocities. It is interesting that the solutions (\ref{eq:exp})\,--\,(\ref{eq:skuma}) do not
depend on the moment of inertia; therefore, the curves representing them for the two limiting cases completely coincide in Fig.~\ref{fig:f1}.
The entire difference between the cases is contained in the values of the wind constant $K_{\rm w}$. These values have to be larger, in proportion to the ratio
$i/i_{\rm e} = 0.074/0.0105\approx 7.05$, for the solid-body rotational evolution because in this case the wind has to be faster
to remove the correspondingly increased amount of angular momentum, so that finally we could get
$\Omega_{\rm e}(t_\odot) = \Omega_\odot$. Note that the last constraint comes directly from the observed rotation period distributions
for solar-type stars in open clusters that show a convergence of $P$ to $P_\odot$ by the solar age.
It is also clear that the correct mechanism of angular momentum redistribution in solar-type stars should not tolerate
a large variation of the wind constant $K_{\rm w}$ between its applications to the fastest and slowest rotators as long as
we believe that the dependence of $\dot{J}$ on $\Omega_{\rm e}$ has correctly been factored out in equation (\ref{eq:jdot}).

The illustrated degeneracy of the solutions (\ref{eq:exp})\,--\,(\ref{eq:skuma}) with respect to the two limiting cases of
coupling between the core and envelope rotation is lifted when we begin to consider more realistic cases in which the coupling is allowed
to increase progressively in time. To demonstrate this via a simple example, let us assume that the moment of inertia of the star's rigidly rotating envelope 
that is decoupled from the rest of it obeys the following rule:
\bea
I(t) = I_{\rm e}\,\left[\frac{\Omega_{\rm e,ZAMS}}{\Omega_{\rm e}(t)}\right]^p,
\label{eq:plaw}
\eea
where $p\equiv\ln(I/I_{\rm e})/\ln(\Omega_{\rm e,ZAMS}/\Omega_\odot)$. Although this case does not have a physical justification, it admits
an analytical solution of equation (\ref{eq:jdot}) and it also guarantees that $I(t_{\rm ZAMS}) = I_{\rm e}$ (no core/envelope coupling
at the beginning) and $I(t_\odot) = I$ (full coupling at the end). 
For $\Omega_{\rm e,ZAMS} = 100\,\Omega_\odot$, the solution of equation (\ref{eq:jdot}) for the dependence (\ref{eq:plaw}) 
is plotted in Fig.~\ref{fig:f1} with a dot-dashed curve.
It needs $K_{\rm w} = 2.56\times 10^{47}$\,cm$^2$\,g\,s. Thus, in spite of the fact that the evolution begins with angular momentum
being lost only from the convective envelope, the wind constant has a value much closer to the previously considered
case of solid-body rotational evolution. This is explained by the fact that the star will later have to get rid of
more angular momentum than in the old decoupled case because the moment of inertia of its decoupled and uniformly rotating envelope
(not just the convective envelope but also an outer part of the radiative core adjoint to it) increases well above $I_{\rm e}$.
Therefore, it is necessary to lose additional angular momentum from the surface of the convective envelope in the future, which will be supplied by
the radiative core as its rotation gets more and more coupled to that of the envelope, demanding a larger wind constant which,
in turn, leads to a steep initial decline of $\Omega_{\rm e}$.

\subsection{The Double-Zone Model}

The double-zone model (\citealt{mcg91}) assumes that the Sun (or a solar-type star) consists of two uniformly rotating zones, the core and envelope, and that
an excess of angular momentum in the core, compared to the case in which the whole star rotates rigidly, is transferred to the envelope on a specified constant timescale
$\tau_{\rm c}$. It is evident that, for this core-envelope rotational coupling to replenish the angular momentum content of the envelope
faster than it is drained by the wind, one needs to have $\tau_{\rm c}\ll\tau_{\rm w} = 133$ Myr (for $\Omega_{\rm e,ZAMS} = 100$), and vice versa.
Computations show (e.g., \citealt{dea09}) that the solid-body rotational evolution of a solar-type star, in the case that its core and envelope rotate nearly
synchronously all the time, can only be achieved with $\tau_{\rm c} \leq 1$ Myr. 

It turns out that the upper 90th percentiles of $\Omega_{\rm e}$ distributions (the fastest rotators) for solar analogs in open clusters
are very well reproduced by the double-zone model with $\tau_{\rm c} = 1$ Myr (upper dotted curves in panels A and B in Fig.~\ref{fig:f2}).
On the contrary, the slowest rotators (the lower 10th percentiles) at ages of order 100 Myr are found to be located below the curves representing the solid-body
rotational evolution ($\tau_{\rm c} = 1$ Myr) that starts from the 10th percentiles of $\Omega_{\rm e}$ distributions for the youngest clusters, even
when the safe upper limit of $\tau_{\rm d} = 20$ Myr is chosen for the disk-locking time (the lower dotted curve in Fig.~\ref{fig:f2}A). 
A rigorous statistical analysis that uses Monte Carlo simulations and compares the full $\Omega_{\rm e}$
distributions (modeled versus observed ones) rather than just their percentiles confirms this conclusion. 
It also gives the following estimates of the coupling time for
the slowest rotators: $\tau_{\rm c}\approx 55\pm 25$ Myr for stars with $0.9\la M/M_\odot\la 1.1$ and
$\tau_{\rm c}\approx 175\pm 25$ Myr for stars with $0.7\la M/M_\odot < 0.9$ (\citealt{dea09}).

\subsection{The Purely Diffusive Model}
\label{sec:difmod}

\begin{figure}
\epsfxsize=14cm
\epsffile [60 270 480 695] {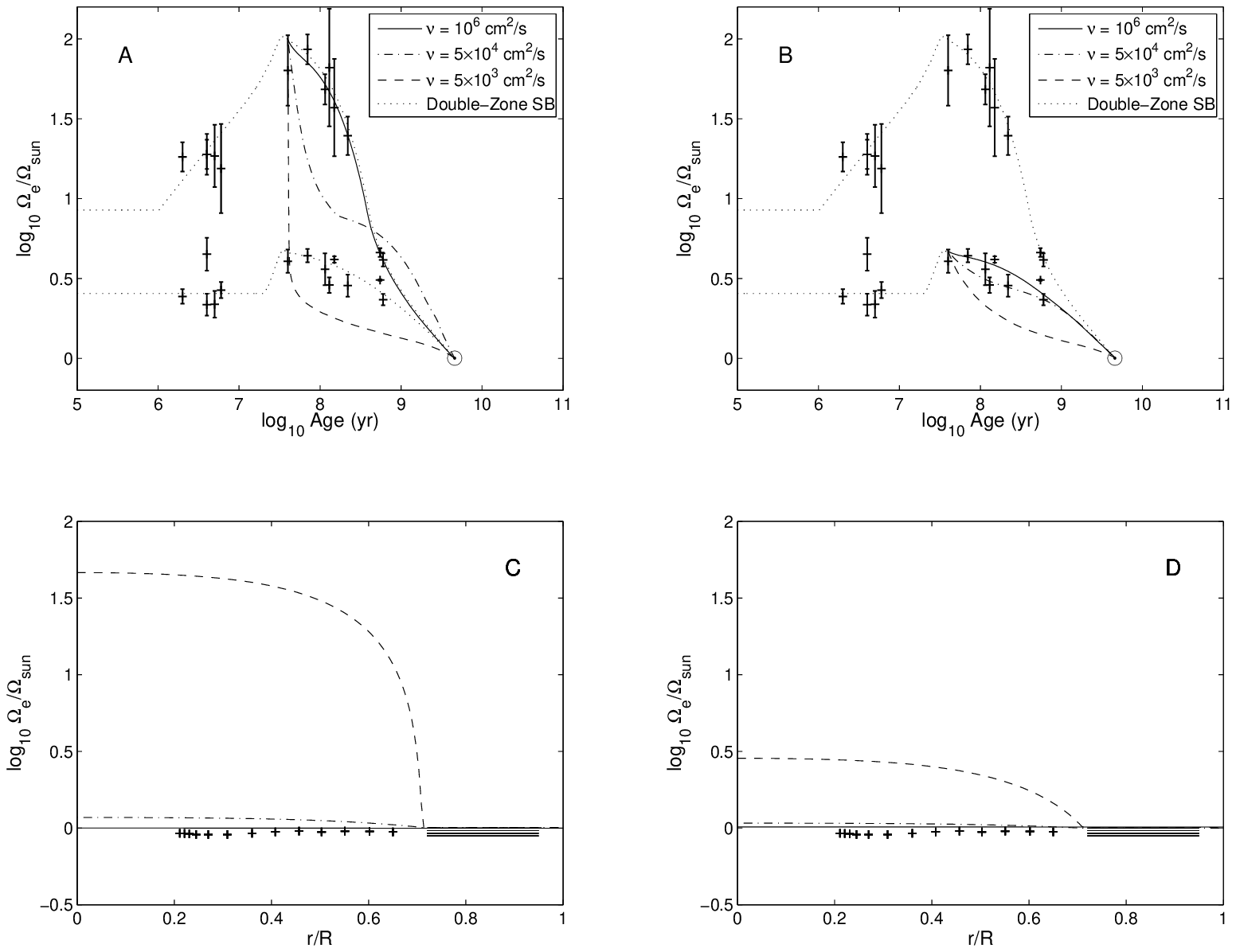}
\caption{Spin-down of the surface solar rotation for the case in which the angular momentum redistribution in the Sun's interior
         is modeled as a purely diffusive process (equation \ref{eq:diff}) with a diffusion coefficient $\nu$
         (displayed in panels A and B for the fastest and slowest rotators, respectively). The corresponding internal rotation profiles
         at the solar age are shown in panels C and D with crosses representing the observational data from
         helioseismology measurements by \cite{cea03}. Horizontal line segments at $r/R\geq 0.713$ in
         the same panels show the latitudinal differential rotation of the (bottom of) Sun's convective envelope at $\theta = 30^\circ$, 
         $45^\circ$, $60^\circ$, and $90^\circ$ (from the lower to upper).
         }
\label{fig:f2}
\end{figure}

The purely diffusive model uses appropriate initial and boundary conditions, including (\ref{eq:jdot}), to solve the following equation:
\bea
\rho r^4\,\frac{\partial\Omega}{\partial t} = \frac{\partial}{\partial r}\left(\rho r^4\nu\,\frac{\partial\Omega}{\partial r}\right),
\label{eq:diff}
\eea
where $\nu$ is a constant viscosity whose physical nature is not specified. The time-dependent density distribution
$\rho(r,t)$ in the radiative core and other necessary stellar structure parameters are taken from full stellar evolution computations.
\cite{dea09} have demonstrated that the solid-body rotational evolution of solar-type stars can only be simulated with
$\nu\ga 10^6$\,cm$^2$\,s$^{-1}$ (solid curves in Fig.~\ref{fig:f2}). They have also established an approximate correspondence between the diffusion coefficient
$\nu$ and the coupling time $\tau_{\rm c}$ from the double-zone model that produce a similar rotational evolution. 
In particular, this relation says that the spin-down of
a slowly rotating star computed using the double-zone model with $\tau_{\rm c}\approx 40$ Myr looks almost identical to that
obtained with the purely diffusive angular momentum redistribution
for $\nu\approx 5\times 10^4$\,cm$^2$\,s$^{-1}$ (dot-dashed curves in Fig.~\ref{fig:f2}). This solution gives an example similar to (\ref{eq:plaw}) when
the moment of inertia of the rigidly rotating envelope effectively increases with time as the diffusive transport of
angular momentum gradually penetrates into the radiative core. It should be noted
that even for a coupling time as long as 90 Myr the corresponding diffusion coefficient still remains
as large as $\nu\approx 2.5\times 10^4$\,cm$^2$\,s$^{-1}$.

The angular momentum redistribution in solar-type stars is known to be accompanied by a chemical element transport, the most
pronounced manifestation of which is the strongly reduced abundance of Li in the solar atmosphere. \cite{mea09} have recently reported
that the Sun does not seem to be unique in this respect. It turns out that solar twins (stars with mass and metallicity
not different from the solar values to within a few percent) show a dependence of the surface Li abundance on
age to which the Sun also belongs (Fig.~\ref{fig:f3}). A natural interpretation of this relation is that
a rate of Li destruction in the atmospheres of solar twins is approximately the same, the Sun simply being a relatively old star.

We have added diffusion terms to a network of nuclear kinetics equations relevant for the Sun's
evolution (the pp chains and CNO cycle reactions). This network has been solved in a post-processing way using,
as a background, previously stored files with all necessary information about the Sun's internal structure from
its fully convective pre-MS evolution to its present age. The resulting time evolution of the surface Li abundance
is plotted in Fig.~\ref{fig:f3} for several values of the diffusion coefficient. It is seen that only
a value of $\nu\approx (2$\,-\,$4)\times 10^3$ cm$^2$\,s$^{-1}$ satisfies the observed anti-correlation.
On the other hand, a value of $\nu = 5\times 10^4$ cm$^2$\,s$^{-1}$ that was used by CHMG93
to model the magnetic braking of solar rotation leads to an excessive Li depletion inconsistent with the observations.
In our modified model of magnetic braking of solar rotation with anisotropic turbulent diffusion, we will
employ a value of $\nu_{\rm v} = 5\times 10^3$\,cm$^2$\,s$^{-1}$ which is close to the values constrained by
the Li abundance. Note that, without being assisted by another angular momentum transport mechanism, the purely
diffusive mixing with this small coefficient can reproduce neither the Sun's nearly uniform rotation nor the spin-down of the fastest rotators             
in open clusters (dashed curves in Fig.~\ref{fig:f2}).

Finally, it is important to note that Be, another fragile element, does not seem to share the fate of Li in evolved MS stars with
close to solar effective temperatures, including the Sun (\citealt{r08}). This means that the chemical mixing
that causes the Li destruction does not penetrate below $r\approx r_{\rm e}-0.16\,R_\odot\approx 0.55\,R_\odot$ in the Sun
(eg., see Fig.~1 of \citealt{bea99}), $r_{\rm e}\approx 0.713\,R_\odot$
being the radius of its core-envelope interface, where the temperature is high enough for protons to begin destroying Be. Alternatively,
it is possible that the diffusion coefficient rapidly declines with depth (e.g., \citealt{rmt00}). If this is true then
the Li mixing may have nothing to do with the global angular momentum redistribution in the Sun. On the other hand, to be in accord with observations
the latter process cannot be assisted or accompanied by a mass transfer whose rate, when described in terms of a diffusion
coefficient, greatly exceeds $\nu\approx 4\times 10^3$\,cm$^2$\,s$^{-1}$.

\section{Dynamic and Stationary Models of Solar Rotation with Large-Scale Magnetic Fields}

\subsection{Magnetic Braking of Solar Rotation: the Old Solution}
\label{sec:chmg93}

Because our main goal is to modify the well-known model of magnetic braking of solar internal differential rotation for it
to enable to explain the recent observational data on the $\Omega_{\rm e}$ distributions for solar-type stars
in open clusters, without causing a conflict with the Li abundance data for solar twins,
we will describe the model employed by CHMG93 that we want to develop only briefly.
For more details about the model, the interested reader is referred to the original paper for an excellent analysis of
numerical results obtained with it and also a discussion of its shortcomings.
To save space, we introduce, at the outset, the general forms of relevant equations that encompass both the original case
considered by CHMG93 and our modified case with anisotropic turbulent diffusion.
In fact, we have chosen an equivalent system of equations similar to that used in the follow-up study by \cite{rk96},
which allows the poloidal field potential to fade with time. 

In the spherical polar coordinates $\{r,\theta,\varphi\}$, 
the magnetic breaking of solar rotation in the presence of anisotropic turbulence in the radiative core is described
by the following equations
\bea \nonumber
\label{eq:mom}
\rho\,r\sin\theta\,\frac{\partial\Omega}{\partial t} & = &
\frac{\sin\theta}{r^3}\frac{\partial}{\partial r}\left(\rho\,r^4\,\nu_{\rm v}\,\frac{\partial\Omega}{\partial r}\right) +
\frac{\rho}{r\sin^2\theta}\frac{\partial}{\partial\theta}\left(\sin^3\theta\,\nu_{\rm h}\,\frac{\partial\Omega}{\partial\theta}\right) \\
& + & \frac{1}{4\pi r^2\,\sin\theta}\left[\frac{1}{r}\,\frac{\partial A}{\partial\theta}\,\frac{\partial (Br)}{\partial r} -
\frac{1}{\sin\theta}\,\frac{\partial A}{\partial r}\,\frac{\partial(B\sin\theta)}{\partial\theta}\right] +
(\rho\,r\sin\theta)\,S_J\Omega, \\ \nonumber \\ \nonumber
\label{eq:ind}
\frac{\partial B}{\partial t} & = & \frac{1}{r}\frac{\partial}{\partial r}\left[\eta_{\rm v}\frac{\partial (Br)}{\partial r}\right] +
\frac{\eta_{\rm h}}{r^2}\frac{\partial}{\partial\theta}\left[\frac{1}{\sin\theta}\frac{\partial (B\sin\theta)}{\partial\theta}\right] \\
& + & \frac{1}{r}\left(\frac{\partial\Omega}{\partial r}\,\frac{\partial A}{\partial\theta} - 
\frac{\partial\Omega}{\partial\theta}\,\frac{\partial A}{\partial r}\right), \\ \nonumber \\
\label{eq:pot}
\frac{\partial A}{\partial t} & = & \eta_{\rm v}\,\frac{\partial^2 A}{\partial r^2} +
\eta_{\rm h}\,\frac{\sin\theta}{r^2}\frac{\partial}{\partial\theta}\left(\frac{1}{\sin\theta}\frac{\partial A}{\partial\theta}\right), \\ \nonumber
\eea
where $S_J = \dot{J}/(I_{\rm e}\Omega_{\rm e})$
is a normalized angular momentum loss rate, with $I_{\rm e}$ and $\Omega_{\rm e}$ being, as before, the moment of inertia and angular velocity of
the convective envelope. For the rate of angular momentum loss from the surface $\dot{J}$, we continue to use equation (\ref{eq:jdot}), whereas CHMG93
used the formulation of \cite{wd67}. 

Equations (\ref{eq:mom})\,--\,(\ref{eq:ind}) represent the $\varphi$\,- (azimuthal) components of the momentum and induction equations.
Following CHMG93, we neglect any meridional motion of either rotational or magnetic origin. We also omit the Coriolis force,
while solving the problem in an inertial (non-rotating) frame of reference in which the polar axis is directed along the rotation axis.
The right-hand side of the momentum equation (\ref{eq:mom}) contains the viscous term, which has been split into its vertical and horizontal components,
the magnetic tension part of the Lorentz force $(1/4\pi)[(\mathbf{B},\nabla)\mathbf{B}]_\varphi$, where 
the total (poloidal plus toroidal) field is $\mathbf{B}\equiv\mathbf{B}_{\rm p} + B\mathbf{e}_\varphi$ 
in which the poloidal field has been expressed via its potential $A$ as follows
\bea
\mathbf{B}_{\rm b} = \nabla\times\left(\frac{A\,\mathbf{e}_\varphi}{r\sin\theta}\right) = 
\left\{\frac{1}{r^2\sin\theta}\,\frac{\partial A}{\partial\theta}, -\frac{1}{r\sin\theta}\,\frac{\partial A}{\partial r}, 0\right\},
\label{eq:BpviaA}
\eea
and, finally, the term simulating the surface angular momentum loss by smearing $\dot{J}$ over the entire
convective envelope, which therefore must vanish in the radiative core. 

Equation (\ref{eq:pot}) is decoupled from the first two equations, therefore it can be solved separately. Following \cite{rk96},
we expand the poloidal field potential into a series of the associated Legendre polynomials and keep only the longest living dipole term
\bea
A(r,\theta,t)\approx -a(r)\sin^2\theta\exp(-t/\tau),
\label{eq:potdip}
\eea
which we substitute into (\ref{eq:pot}). As a result, we have to solve the eigenvalue problem
\bea
\eta_{\rm v}\,\frac{d^2a}{dr^2} - 2\,\frac{\eta_{\rm h}}{r^2}\,a = -\frac{a}{\tau}
\label{eq:eig}
\eea
for the boundary conditions $a(0) = a(r_{\rm b}) = 0$, where $0\leq r_{\rm b}\leq R_\odot$, and the potential $A$ will thus be determined.

CHMG93 considered the isotropic case in which $\nu\equiv\nu_{\rm v}\equiv\nu_{\rm h}$ and $\eta\equiv\eta_{\rm v}\equiv\eta_{\rm h}$.
They artificially amplified the microscopic kinematic viscosity $\nu$ and magnetic diffusivity $\eta$ by factors
of order $10^3$ and $10^2$, respectively, making them as large as $\nu = 5\times 10^4$\,cm$^2$\,s$^{-1}$ and $\eta = 5\times 10^5$\,cm$^2$\,s$^{-1}$
and keeping their values constant throughout the radiative core. It is convenient to transform equation (\ref{eq:eig}) into its dimensionless form
\bea
\frac{d^2a}{dy^2} - \frac{2}{y^2}\,\gamma\,a = -\lambda\,a,
\label{eq:eigdim}
\eea
where $y\equiv r/r_{\rm b}$, $\gamma\equiv\eta_{\rm h}/\eta_{\rm v}$, and $\lambda\equiv (r_{\rm b}^2/\tau\eta_{\rm v})$.
Taking the values of $\gamma = 1$ and $\eta_{\rm v} = 5\times 10^5$\,cm$^2$\,s$^{-1}$ used by CHMG93, we
solve the eigenvalue problem (\ref{eq:eig}\,--\,\ref{eq:eigdim}) and find that, in this case, the poloidal field potential can be
approximated by the following relation:
\bea
A(r,\theta,t)\approx B_0\,\frac{r^2}{2}\left(1-\frac{r}{r_{\rm b}}\right)^2\sin^2\theta\times \exp(-t/\tau),
\label{eq:potential}
\eea
where $\tau = 35$ Myr for $r_{\rm b} = r_{\rm e} = 0.713\,R_\odot$. The same, or very similar, form of the potential was used by both CHMG93 and \cite{rk97}.
However, in both of those investigations the time dependence of $A$ was ignored, which might not be a good assumption,
given the short decay time for $A$ obtained with that for the enhanced magnetic diffusivity. 

The decay time can be as high as 5.4 Gyr, which is comparable
to the solar age, only when the maximum value of the microscopic magnetic diffusivity in the Sun's radiative
core $\eta_{\rm max}=\eta(r_{\rm e})\approx 3\times 10^3$\,cm$^2$\,s$^{-1}$ is substituted into the above expression for 
$\lambda$ instead of $\eta_{\rm v}$ (cf. \citealt{rk96}). But, in this case, magnetic braking
will hardly work, especially, after our having correspondingly decreased the amplified macroscopic viscosity either to the maximum microscopic
value of $\nu_{\rm max} = \nu(r_{\rm e})\approx 20$\,cm$^2$\,s$^{-1}$ or to its minimum turbulent value of $\nu\approx\eta =\eta_{\rm max}$
(assuming that $\eta$ is still of turbulent origin but has also reached its minimum possible value).
The latter substitution brings the turbulent magnetic Prandtl number $P_{\rm m} \equiv \eta/\nu$
close to its most plausible value of unity.                    

Keeping this inconsistency in mind, we proceed with
our revision of the original model of CHMG93. Like them, we apply equations (\ref{eq:mom})\,--\,(\ref{eq:ind}) and
(\ref{eq:potential}) to the whole star assuming that $\nu$ and $\eta$ jump to their large convective turbulent values of order
$10^{12}$\,cm$^2$\,s$^{-1}$ at the core-envelope interface. Solutions for the angular velocity $\Omega(r,\theta,t)$ and
toroidal magnetic field $B(r,\theta,t)$ are sought in the space domain $(r,\theta) = [0,R_\odot]\times [0,\pi]$
on the time interval $t_{\rm ZAMS}\leq t\leq t_\odot$. Following CHMG93, it is convenient to use $\mu\equiv\cos\theta$
as the second independent variable instead of $\theta$, in which case the relevant space domain turns into the rectangular
area $(x,\mu) = [0,1]\times [-1,1]$, where $x=r/R_\odot$. 

We start our computations with the following initial conditions:
$\Omega(r,\theta,0) = \Omega_{\rm e,ZAMS}$, and $B(r,\theta,0) = 0$. The model's internal structure is considered to be fixed
and described by the density distribution $\rho(r,t_\odot)$ for the present-day solar model taken from our full stellar
evolution computations (\citealt{dea09}). For the dimensionless moment of inertia of the convective envelope, we
use the time-averaged value of $i_{\rm e} = 0.0105$. Given that the surface loss of angular momentum is already included
in equation (\ref{eq:mom}) and that the equations are solved for the whole star, the boundary conditions become simple:
$\partial\Omega/\partial r(R_\odot,\theta,t) = 0$, $\partial\Omega/\partial\theta = 0$ at each point on the rotation axis,
and $B = 0$ on all the boundaries, including the core-envelope interface, the latter condition being imposed because
of the extremely large turbulent magnetic diffusivity in the convective envelope that makes it feel like a vacuum to
the internal magnetic field (\citealt{rk96}).

We consider two poloidal field configurations of the four cases investigated by CHMG93. They are specified by
the parameter $r_{\rm b}$ that defines the radial extent of non-vanishing potential $A$ with respect to $r_{\rm e}$
(equation \ref{eq:potential}). We have chosen the qualitatively different configurations
D3 and D2 in the notations of CHMG93. In the most promising, according to CHMG93 and \cite{mgch99}, configuration D3 the potential $A$
occupies the whole radiative core up to the bottom of convective envelope, i.e. we have $r_{\rm b} = r_{\rm e}$ 
(Fig.~\ref{fig:f5}A). The configuration D2 is set up by using the value of $r_{\rm b} = 0.8\,R_\odot > r_{\rm e}$,
which means that $A$ can now penetrate into the lower part of the convective envelope (Fig.~\ref{fig:f5}B) but, 
what is more important, the poloidal field has now a non-vanishing radial component at $r=r_{\rm e}$ for some $\theta$.
Given that $\mathbf{B}_{\rm p}$ in the configuration D2 is thus directly coupled to the envelope rotation, it can
more effectively transfer into the radiative core the torque applied to the convective envelope by the magnetized wind.
In the D3 case, there is a time delay before envelope braking will be felt in the core. This time is taken for the viscous
transport of angular momentum to build up a sufficiently strong rotational shear just beneath the core-envelope interface which will then begin to interact
with $\mathbf{B}_{\rm p}$. For a comparison, we also introduce a purely diffusive configuration D0 with $A\equiv 0$
(Fig.~\ref{fig:f7}A). Finally, our modified magnetic braking models with anisotropic turbulent diffusion that have the D3 and D2
poloidal fields and strong horizontal components of turbulent viscosity in the radiative core will
be referred to as D3H and D2H.

To numerically solve the partial differential equations (\ref{eq:mom})\,--\,(\ref{eq:ind}) in the 2D space domain
with the potential $A$ given by equation (\ref{eq:potential})
we make use of the COMSOL Multiphysics software package. Like the custom-made computer code employed by CHMG93, the COMSOL Multiphysics
solves PDEs using the finite element method. In our computations, we use quadrilateral mesh elements to partition the space domain into a total number of
elements that slightly varies from one solution to another but that is approximately equal to $100\times 200$ and comparable
to the mesh sizes used by CHMG93. Note that we seek a solution in a half meridian, whereas CHMG93 did their computations for a quadrant.
This reduces the resolution of our mesh by a factor of two compared to theirs.
Otherwise, apart from the different prescription for the surface loss of angular momentum
and the option of taking into account the exponential decay of $\mathbf{B}_{\rm p}$, our analysis of
magnetic braking of solar rotation with the isotropic viscosity and magnetic diffusivity is very similar to that by CHMG93.

Analyzing the results of their extensive computations, CHMG93 have come to the conclusion that the rotational evolution of
the Sun weakly depends on the poloidal field amplitude $B_0$ (equations \ref{eq:BpviaA} and \ref{eq:potential}).
On the contrary, the average strength of the toroidal field that is generated by differential rotation in the radiative core changes significantly with $B_0$,
a weaker poloidal field resulting in a stronger toroidal field. This anti-correlation emerges from the necessity of the magnetic torque to compensate
for the one exerted by the magnetized solar wind, the former being proportional to the products $B_r B_\varphi$ and $B_\theta B_\varphi$ in which 
$B_r\propto B_0$ and $B_\theta\propto B_0$ because the total magnetic field in the radiative core is
$\mathbf{B} = \mathbf{B}_{\rm p} + B\,\mathbf{e}_\varphi$.

CHMG93 have distinguished the following three important epochs in the magnetic braking of solar rotation that differ from each other by
the relative strength of the magnetic and wind torques. During the first epoch, the magnetic torque is much weaker than the wind torque and,
as a consequence, the rotational shear produced by the spin-down of the convective envelope at its interface with the radiative core
is spread by the viscosity almost without hindrance to the core surface. So, the shear in the outer core grows almost linearly with time and, as a result,
the toroidal field is amplified there almost quadratically in time. This short epoch, lasting between a few thousand to a few million years,
is followed by the second epoch during which time the differential rotation profile begins to interact with the growing toroidal field.
For the poloidal field configurations in which the core and envelope are magnetically coupled (e.g., the D2 case), this epoch is characterized
by vigorous phase mixing throughout the radiative core. This process is initiated by large scale toroidal field oscillations that originate
close to the bottom of convective envelope and propagate through the radiative core
along the poloidal field lines using them as elastic rubber strings. Because the strings are rooted in the
convective envelope (for the D2 configuration) at different colatitudes and therefore have different lengths, 
the Alfv\'{e}n waves carrying the perturbations back and forward along the strings get out of phase
as time progresses. Eventually, this causes large gradients of toroidal magnetic field to build up locally at different places and, as a result,
the large-scale toroidal field oscillations are diffused away either via ohmic dissipation or via turbulent magnetic 
reconnection, depending on the nature of $\eta$ (e.g., \citealt{s99}).
Finally, during the third epoch, a state of dynamical balance that is achieved at the end of the second epoch
between the magnetic torque in the radiative core and the total stresses
(magnetic plus viscous) at the bottom of convective envelope is maintained 
as the surface wind torque decreases with time.

First, we have repeated the computations of CHMG93 for one of their most favorite models, 
namely the one that starts with $\Omega_{\rm e,ZAMS} = 50\,\Omega_\odot$
and has the D3 configuration of time-independent poloidal field with
the amplitude $B_0 = 1$\,G. Our results are presented in Fig.~\ref{fig:f4} (thick solid curves in panels A and C as well as panels B and D). 
The geometry and magnitude of the residual toroidal field in
the present-day Sun (panel D) look very similar to those obtained by CHMG93 (upper quadrant of the lower right panel in their Fig.~6).
As they reported, the model has a nearly flat rotation profile by the solar age (panels B and C). However, it seems unlikely that it will
be able to fit the upper 90th percentiles of the $\Omega_{\rm e}$ distributions for open-cluster solar analogs after its initial rotation rate
is increased to a value of $\Omega_{\rm e,ZAMS} = 100\,\Omega_\odot$ corresponding to the observed maximum (the peak on the dotted curve
representing the solid-body rotational evolution simulated using the double-zone model). This supposition is confirmed by our additional
test computations that started with $\Omega_{\rm e,ZAMS} = 100\,\Omega_\odot$ (thin solid curve in panel A). 

It is also not clear what could possibly force
the rotational evolution of the CHMG93 model to change its morphology in the $\Omega_{\rm e}$ vs. Age plane
from the convex one for the fastest rotators to the concave one for the slowest rotators,
as is indicated by both direct observations (e.g., compare the solid and dot-dashed curves in panels A and B in Fig.~\ref{fig:f2}, respectively)
and their rigorous statistical analysis (\citealt{dea09}).
Physically, the change implies a transition from solid-body rotational evolution to evolution with a differential rotation
sustained for a few tens to a hundred Myr. As has been mentioned, this cannot be produced by a variation of $B_0$. We can see now that
an increase of the initial rotation rate does not make the rotational evolution of the CHMG93 model approach the dotted curve either.
In other words, this model does not have enough internal degrees of freedom to explain the whole spectrum of observed rotation rates.
Besides, its high viscosity ($\nu = 5\times 10^4$\,cm$^2$\,s$^{-1}$) causes the model to conflict with 
the observed Li abundances in solar twins (Section \ref{sec:difmod}).
When we reduce the viscosity to the value of $\nu = 5\times 10^3$\,cm$^2$\,s$^{-1}$, which is more or less compatible with the Li data,
then the model's rotational evolution becomes absolutely incapable of reproducing the upper 90th percentiles of the $\Omega_{\rm e}$
distributions (dashed curve in Fig.~\ref{fig:f4}A).

As far as the importance of taking into account the exponential decay of the poloidal field is concerned, it depends on how big the values of
$\gamma = \eta_{\rm h}/\eta_{\rm v}$ and $\eta_{\rm v}$ are in the model in question. For the above considered case of 
$\Omega_{\rm e,ZAMS} = 100\,\Omega_\odot$, $\gamma = 1$ and 
$\eta_{\rm v} = 5\times 10^5$\,cm$^2$\,s$^{-1}$, a more consistent solution with $A$ decaying on the timescale of $\tau = 35$ Myr
(equation \ref{eq:potential}) is plotted in Fig.~\ref{fig:f4}A as a dot-dashed curve. Although it does not appear to strongly deviate
from the old solution (the thin solid curve), the difference between the two solutions will probably be significant for Monte Carlo simulations and a rigorous
statistical analysis of rotation period distributions for solar analogs in open clusters like those performed by \cite{dea09}.

Finally, it is worth noting that the purely diffusive model with the same viscosity $\nu = 5\times 10^4$\,cm$^2$\,s$^{-1}$ as
the one used by CHMG93 seems to produce both the Sun's rotational evolution and its final internal rotation profile (the dot-dashed curves in panels A and C
in Fig.~\ref{fig:f2}) that are not very different from those computed with the model of CHMG93, especially
when the poloidal field potential is allowed to decrease with time (the dot-dashed curves in panels A and C in Fig.~\ref{fig:f4}).  
The secondary role of magnetic braking compared to the viscous transport of angular momentum is emphasized when we take a much smaller value of
$\nu = 5\times 10^3$\,cm$^2$\,s$^{-1}$, consistent with the solar Li data (compare dashed curves in the same four panels).
In this case, the only noticeable effect produced by the presence of magnetic fields is a slightly slower rotation of
the radiative core at the solar age, especially close to its center, which still strongly deviates from the helioseismology
data.

\subsection{Magnetic Braking With Anisotropic Turbulent Diffusion}
\label{sec:anisotrop}

We start with the aforementioned failed D3 model of magnetic braking of solar rotation that has 
an isotropic viscosity $\nu = 5\times 10^3$\,cm$^2$\,s$^{-1}$ and magnetic diffusivity
$\eta = 5\times 10^4$\,cm$^2$\,s$^{-1}$ and which begins its rotational MS evolution with $\Omega_{\rm e,ZAMS} = 100\,\Omega_\odot$
(thin dashed curves in Fig.~\ref{fig:f6}). Note that its viscosity is already increased by nearly two orders of magnitude compared to
the maximum value of the microscopic (molecular for the Sun) viscosity in the radiative core. 
This is not a problem though, given that vertical diffusion with a coefficient of a similar or slightly stronger magnitude is indeed predicted      
to occur in radiative zones of solar-metallicity MS stars with $M = 1.5\,M_\odot$, especially close to the ZAMS (\citealt{pea03}), where
internal gradients of the mean molecular weight are too weak to effectively hinder it. This diffusion is associated with either rotation-driven
meridional circulation or small-scale turbulence caused by rotational shear instabilities. It has been successfully used to explain  
chemical composition anomalies in the atmospheres of massive MS stars and their descendants (e.g., \citealt{mm00,d05}).

\begin{figure}
\epsscale{0.85}
\plotone{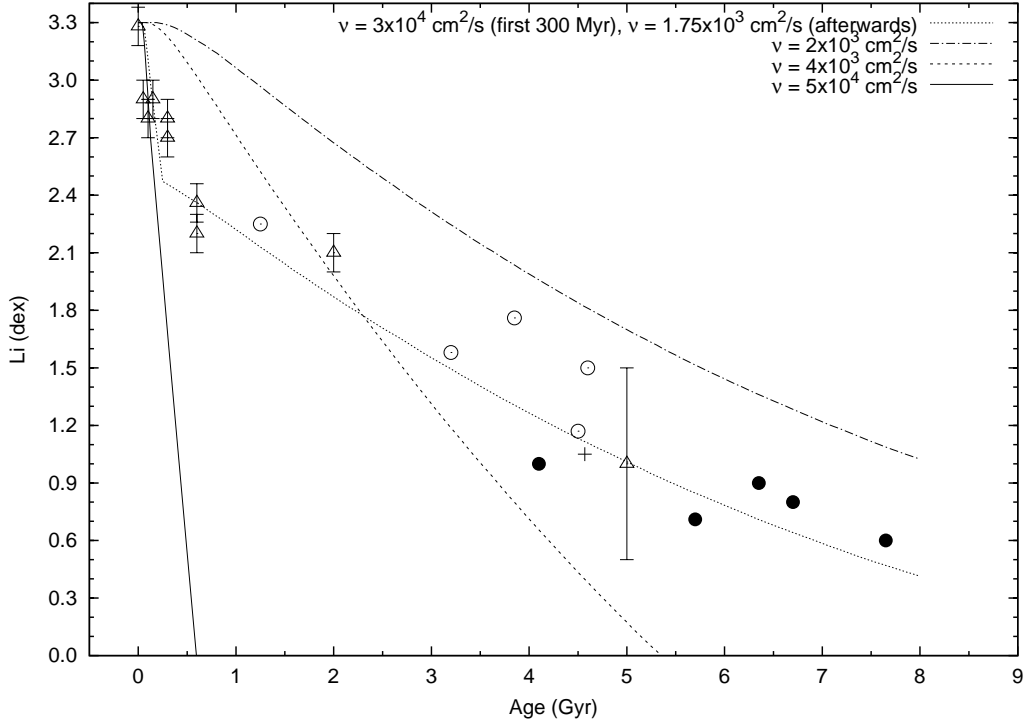}
\caption{Evolutionary decline of the atmospheric Li abundance in solar twins. Symbols represent the observational data of
         \cite{mea09}. Curves are our solar model post-processing computations of the kinetics of nucleosynthesis in the pp chains
         and CNO cycle with chemical mixing modeled as a purely diffusive process. For Li to start being destroyed, it has to be mixed
         at least down to $r\approx 0.65\,R_\odot$ (in the present-day Sun, the core-envelope interface is located at $r=r_{\rm e}\approx 0.713\,R_\odot$).
         For mixing penetrating below $r\approx 0.6\,R_\odot$, the surface Li depletion depends almost entirely on
         the value of the diffusion coefficient $\nu$ and not on the mixing depth (curves). The dotted curve represents the possibility of
         $\nu$ being enhanced during the first 300 Myr.
         }
\label{fig:f3}
\end{figure}

\begin{figure}
\epsfxsize=11cm
\epsffile [10 120 480 695] {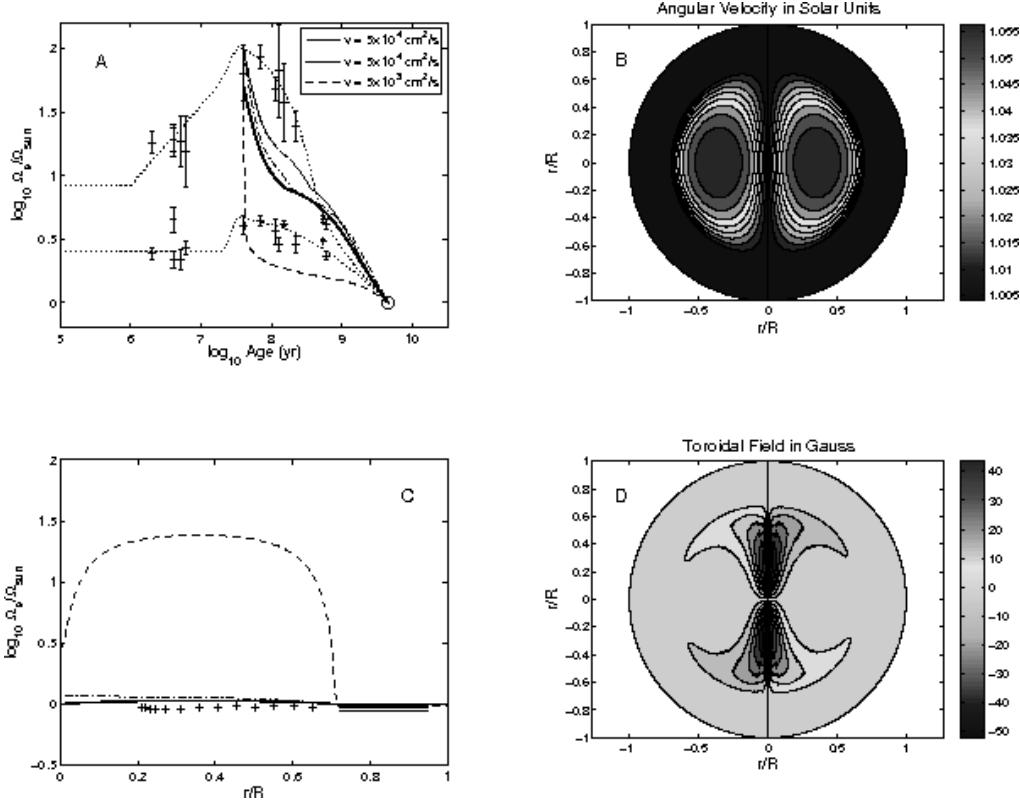}
\caption{Magnetic braking of solar rotation computed using the D3 poloidal field configuration (Fig.~\ref{fig:f5}A)
         with the isotropic viscosity $\nu = 5\times 10^4$\,cm$^2$\,s$^{-1}$ and magnetic diffusivity $\eta = 5\times 10^5$\,cm$^2$\,s$^{-1}$ 
         (thick solid curves in panels A and C as well as contours of constant angular velocity and toroidal field
         in panels B and D, the last three panels showing solutions at the solar age). 
         These values are equal to those used in the favorite model of CHMG93. 
         In the models represented by dot-dashed, thin solid and dashed curves, the initial angular velocity was increased from
         $50\,\Omega_\odot$ to $100\,\Omega_\odot$, and in the first of them the poloidal field of the same amplitude $B_0 = 1$ G as in CHMG93 was allowed to decay 
         with time (equations \ref{eq:eigdim}\,--\,\ref{eq:potential}). The CHMG93-type solution obtained with the reduced values of 
         $\eta = 5\times 10^4$\,cm$^2$\,s$^{-1}$ and $\nu = 5\times 10^3$\,cm$^2$\,s$^{-1}$, the latter being more or less compatible with 
         the observed Li depletion in solar twins, fails to reproduce both the spin-down of the fastest rotating solar-type stars in open 
         clusters and the internal rotation profile of the present-day Sun (dashed curves in panels A and C).
         }
\label{fig:f4}
\end{figure}

Whereas the model of magnetic angular momentum transport based on the hypothetical Tayler-Spruit dynamo has a problem reproducing
differential rotation in the slowest rotators (\citealt{dea09}), the D3 model of CHMG93, on the contrary, cannot provide
a rapidly rotating solar-type star with a means to spin-down its radiative core fast enough for the star's rotational evolution
to fit the upper 90th percentiles of the $\Omega_{\rm e}$ distributions for solar analogs in open clusters. The question is whether
we can make a reasonable modification of the CHMG93 model such that it will become capable of reproducing the rotational evolution of
both the fastest and slowest rotators while being consistent with the other observational constraints? We have shown that a variation of
the isotropic viscosity does not help to solve the problem self-consistently with or without a magnetic field. The major obstacles
are the observational limit of $\nu \la 5\times 10^3$\,cm$^2$\,s$^{-1}$ imposed by the rate of the evolutionary Li depletion
in solar twins (Fig.~\ref{fig:f3}) and the rapid decay of the poloidal field caused by high viscosity and magnetic diffusivity. 
On the theoretical side, stellar hydrodynamics does not predict a build-up of vertical turbulent viscosities
as large as $\nu\sim 10^6$\,cm$^2$\,s$^{-1}$ in the radiative cores of rapidly spinning MS solar-type stars, as required for the stars
to evolve like solid-body rotators. Besides, such large diffusion coefficients would solve the problem without being
assisted by magnetic fields. 

\begin{figure}
\epsfxsize=11cm
\epsffile [20 50 450 695] {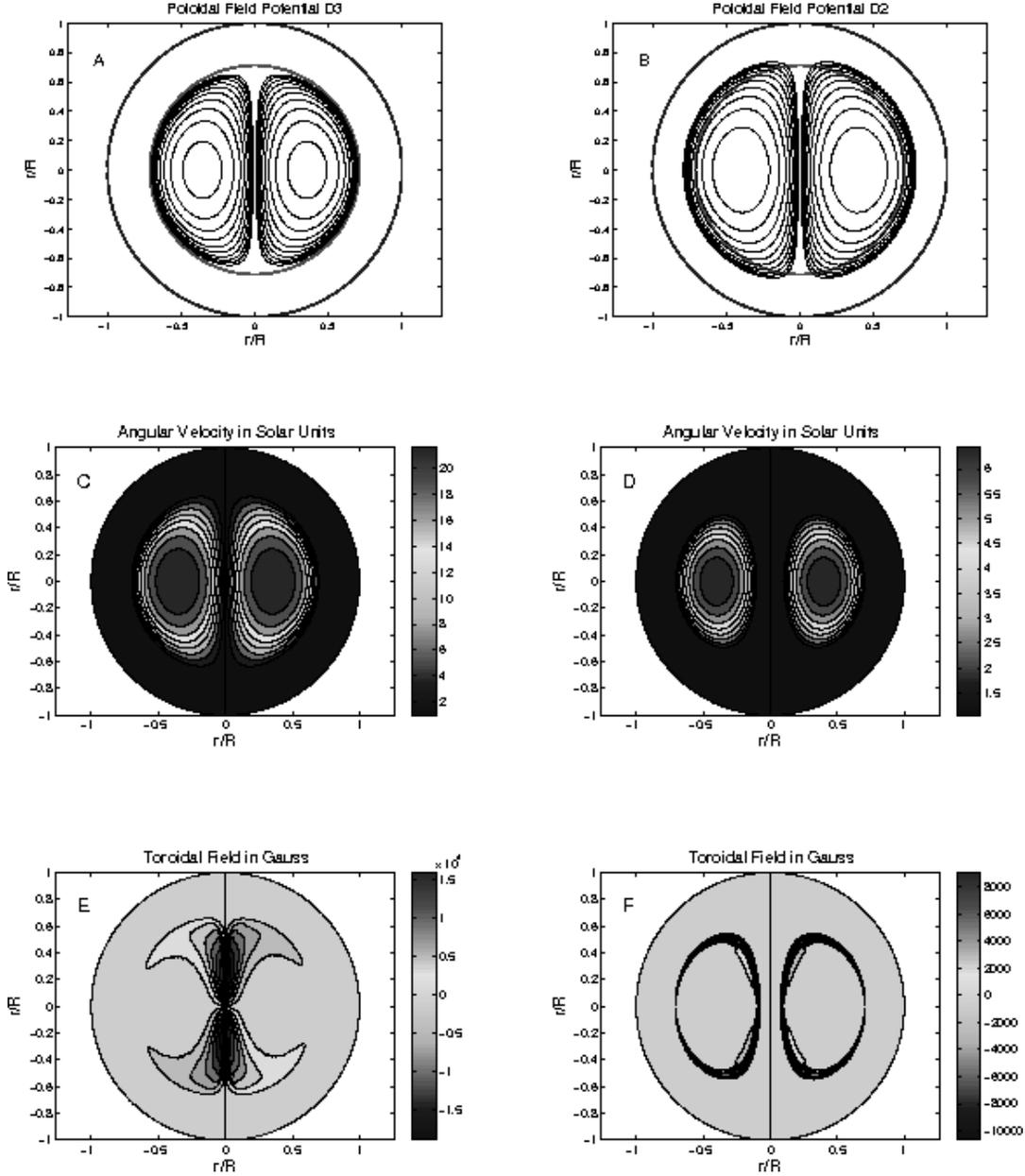}
\caption{Geometries of the poloidal field potential, that is assumed to have here the constant amplitude $B_0 = 0.1$ G, in the configurations 
         D3 (panel A) and D2 (panel B), and their corresponding final 
         (solar age) contours of constant angular velocity and toroidal field (panels C and E, and D and F, respectively) obtained
         in our computations of magnetic braking of solar rotation with the isotropic values of $\nu = 5\times 10^3$\,cm$^2$\,s$^{-1}$ and 
         $\eta = 5\times 10^4$\,cm$^2$\,s$^{-1}$. Note the strong residual differential rotation in the torus-shaped ``dead zones'' (panels C and D).
         Red circles in panels A and B show the core-envelope interface.
         }
\label{fig:f5}
\end{figure}

\begin{figure}
\epsfxsize=14cm
\epsffile [40 270 480 695] {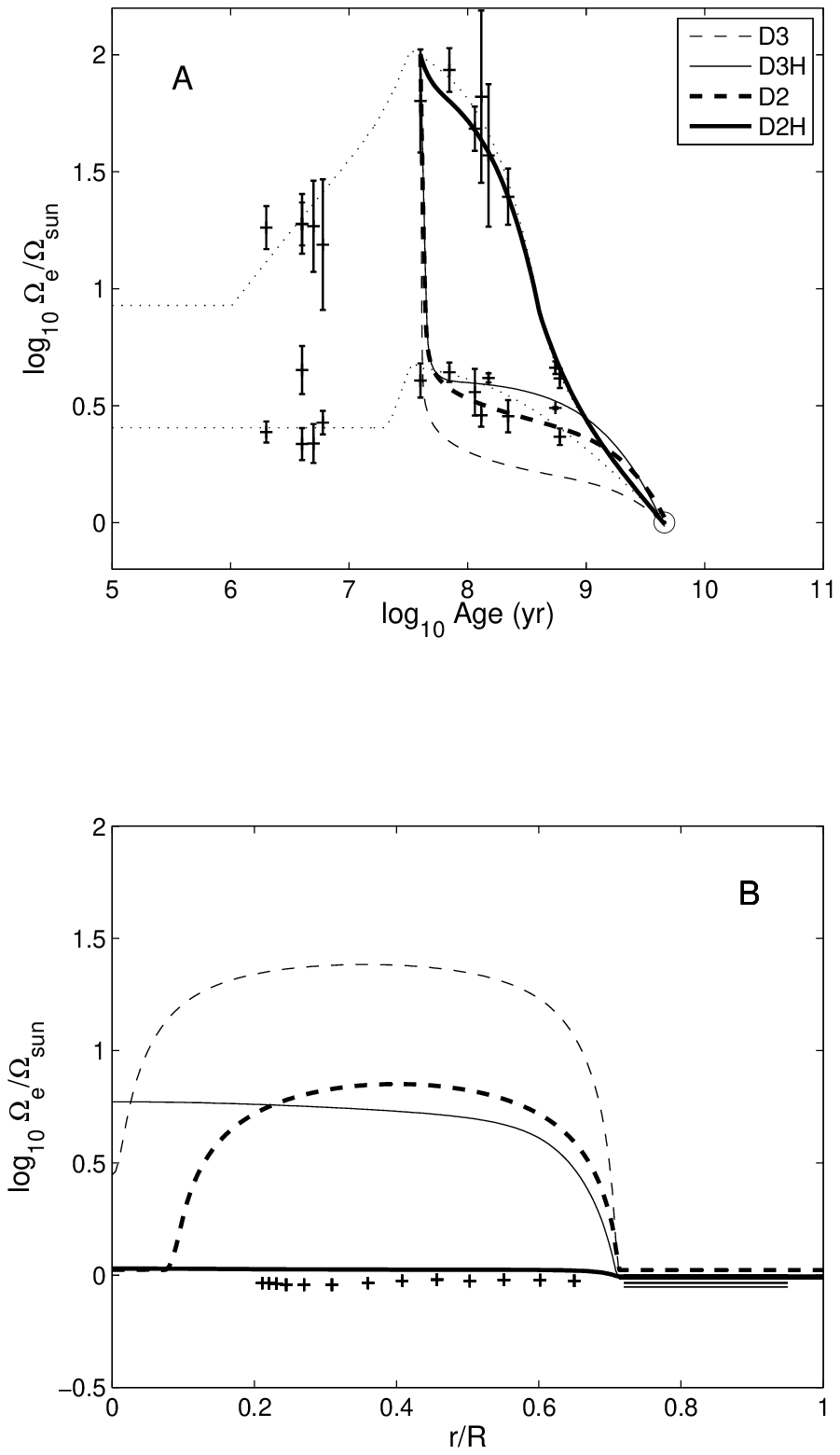}
\caption{Comparison of four similar cases of magnetic braking of solar rotation for the poloidal field configurations D3 and D2 with (suffix ``H'') and without
         a strong horizontal turbulent diffusion ($\nu_{\rm h}\gg\nu_{\rm v}$). In the cases D2H and D3H,
         $\nu_{\rm v} = 5\times 10^3$\,cm$^2$\,s$^{-1}$, $\eta_{\rm v} = 2.5\times 10^3$\,cm$^2$\,s$^{-1}$, and the poloidal field of the amplitude $B_0 = 0.1$ G is
         allowed to decay.  In the D2 and D3 cases, we used $\nu = 5\times 10^3$\,cm$^2$\,s$^{-1}$, $\eta = 5\times 10^4$\,cm$^2$\,s$^{-1}$, 
         and a constant poloidal field of the same amplitude. In the D2H and D3H cases, we assumed that $\nu_{\rm h} = \eta_{\rm h} = 10^6$\,cm$^2$\,s$^{-1}$.
         Panel B compares the final internal rotation profiles with the helioseismology data. Note that only our D2H model gives a satisfactory solution
         for the fastest rotators (thick solid curves).
         }
\label{fig:f6}
\end{figure}

\begin{figure}
\epsfxsize=12cm
\epsffile [10 120 480 695] {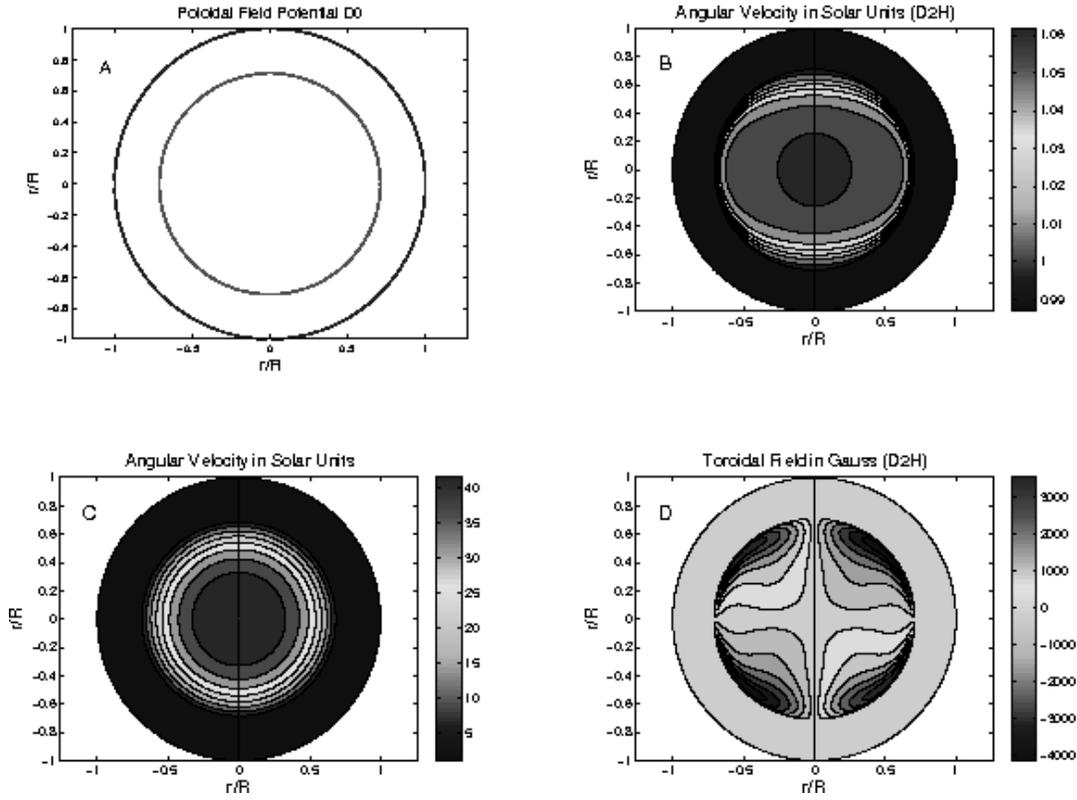}
\caption{Panel A: a sketch of the D0 model of a solar-type star without magnetic fields, the red circle representing the bottom of its convective envelope.
         Panels B and C: contours of constant angular velocity in our D2H model and in the non-magnetic model at the solar age. Panel D: contours of constant
         toroidal magnetic field in the final D2H model.
         }
\label{fig:f7}
\end{figure}

It should be noted that the poloidal magnetic field does help the viscous transport to reduce, even a little, the degree of differential rotation
in the Sun's radiative core, as claimed by CHMG93. Indeed, panels C in our Figs.~\ref{fig:f7} and \ref{fig:f5} indicate that, whereas
the contours of constant angular velocity in the present-day Sun's D0 model are concentric spheres, 
they become consecutive layers in meridian cross sections of an axisymmetric torus
in the D3 model. This torus, called the ``dead zone'' by \cite{mw87}, is shaped by the poloidal field, 
whose lines form a similar toroidal structure (Fig.~\ref{fig:f5}A), according to Ferraro's law
$(\mathbf{B}_{\rm p},\nabla)\Omega = 0$. Thus, in the D3 configuration, the viscous transport has to redistribute angular momentum throughout
a smaller volume than in the absence of magnetic fields (the D0 case). However, when $\nu = 5\times 10^3$\,cm$^2$\,s$^{-1}$,
it does not have enough time to finish its work by the solar age. As CHMG93 reported, the situation is slightly improved when the poloidal field configuration
is switched to D2. In this case, there is no waiting time before the spin-down of the convective envelope gets communicated to
the radiative core because poloidal field lines are now rooted in the envelope, hence they can start transferring the wind torque to
the core from the very beginning. Besides, because not every poloidal field line is now parallel to the bottom of convective envelope
(some of them emanate from it, being directed inside the core), a larger cylindrical region around the rotation axis (not just the axis
and its immediate adjacent layers as it was in the D3 case) acquires a rotation rate close to that of the envelope. As a result,
the dead zone becomes thinner (Fig.~\ref{fig:f5}D). However, even in this case the rotational evolution is still far from being close to that of
a solid-body rotator (thick dashed curves in Fig.~\ref{fig:f6}).

When analyzing the above results, especially contour plots C and D in Fig.~\ref{fig:f5}, we have noticed that
a strong horizontal turbulence, a concept with different approximate descriptions that have already been used in other stellar evolution computations, 
might solve the problem of magnetic braking of solar rotation as well. In fact, the same hydrodynamic models that
give an estimate of $\nu_{\rm v}\sim 10^2$\,--\,$10^4$\,cm$^2$\,s$^{-1}$ for the vertical component of turbulent diffusion also
predict a much stronger value, of order $10^4$\,--\,$10^6$\,cm$^2$\,s$^{-1}$, for its horizontal component in the radiative core of 
the solar-metallicity MS star with $M = 1.5\,M_\odot$ and $\Omega_{\rm e,ZAMS}\approx 50\,\Omega_\odot$ (\citealt{pea03}).
In a solar-mass star the corresponding turbulent viscosities are expected to be lower by a factor of 10, which is roughly proportional to the ratio of luminosities
for MS stars with $M = 1.5\,M_\odot$ and $M = M_\odot$, because
the baroclinic and secular Kelvin-Helmholtz instabilities driving this anisotropic turbulence have growth rates inversely proportional to the thermal timescale.
On the other hand, the coefficient of horizontal turbulent diffusion has been suggested to take on much greater values (by more than four orders of
magnitude) than previously thought (cf. \citealt{z92,m03,mea07}). 

\cite{z92} was the first to bring anisotropic turbulent diffusion
to the attention of the stellar astrophysics community. Anisotropic turbulence has both quite a natural 
physical justification and an experimental basis in geophysics and oceanography.
\cite{t00} notes that ``in the Earth's lower atmosphere one has $\nu_{\rm h}/\nu_{\rm v}\la 10^2$, whereas this ratio may be as large as
$10^5$ in the surface layer of the ocean where large-scale currents are observed''. It is even more likely to occur in stellar radiative zones
with their extremely large Reynolds numbers and strong thermal stratifications in the vertical direction.

\begin{figure}
\epsfxsize=15cm
\epsffile [40 270 480 695] {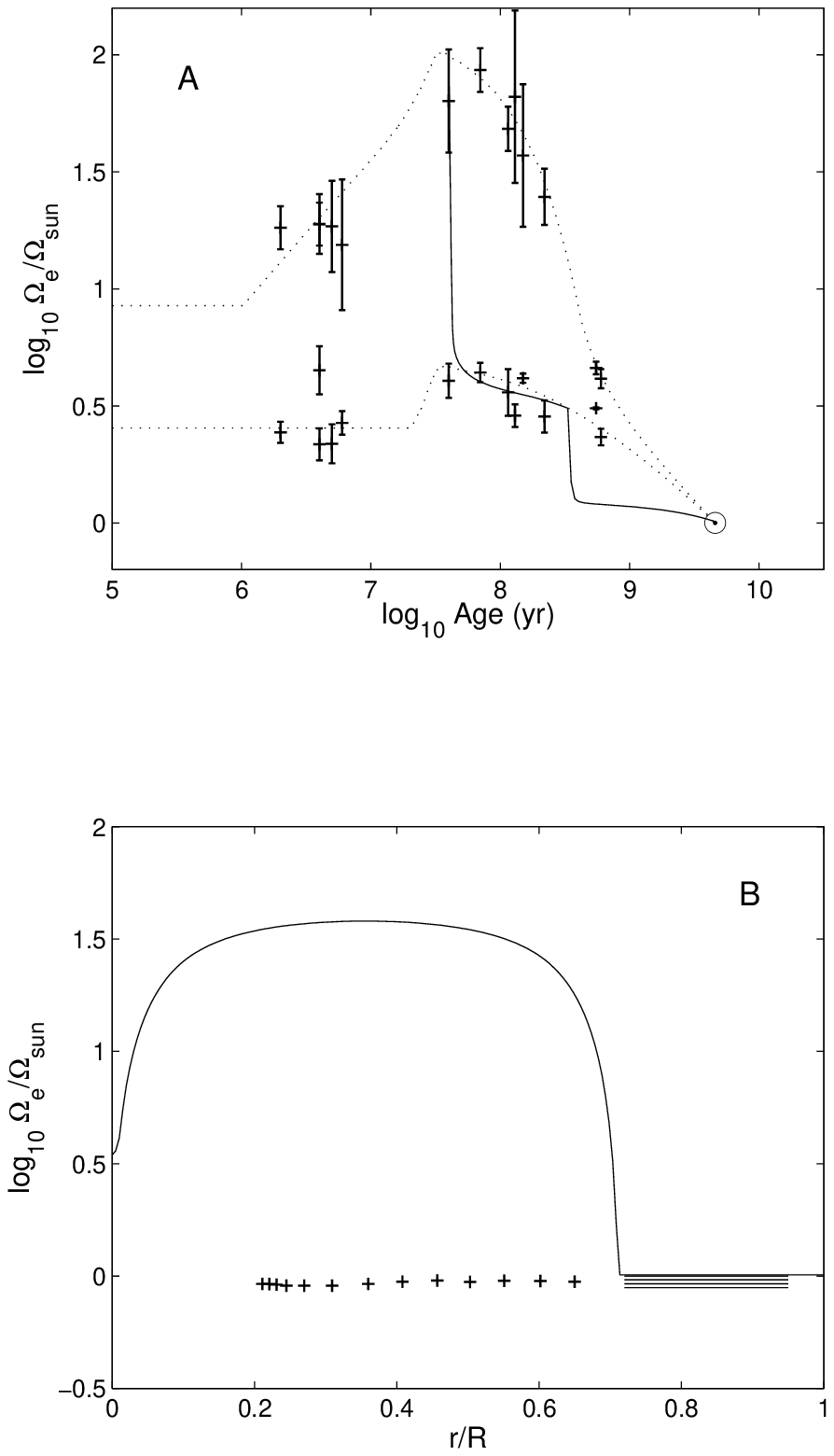}
\caption{Illustration that the assumption of $\nu$ being enhanced during the first 300 Myr (dotted curve in Fig.~\ref{fig:f3})
         does not help to solve the problem using the D3 poloidal field configuration.
         }
\label{fig:f8}
\end{figure}

Unfortunately, neither $\nu_{\rm v}$ nor $\nu_{\rm h}$ can be derived from first principles yet. Therefore, crude estimates are usually adopted 
and then tuned up
by making an order of magnitude evaluation of the relevant energy balance, using appropriate results from laboratory or numerical experiments, and, last
but not least, trying to comply with available observational constraints concerning mixing in stars. Given this uncertainty, while pursuing our modest goal of
producing a modified CHMG93 model that should be as simple as the original one, we have chosen the following prescriptions for
the anisotropic diffusion coefficients in solar-type MS stars: 
\bea
\nu_{\rm v} = 5\times 10^3\,\mbox{cm}^2\,\mbox{s}^{-1},\,\ \ \mbox{and}\ \ 
\nu_{\rm h} = 10^6\,\left(\frac{\Omega_{\rm e,ZAMS}}{100\,\Omega_\odot}\right)\,\mbox{cm}^2\,\mbox{s}^{-1}.
\label{eq:nuvnuh}
\eea
The first expression represents the approximate upper limit for the diffusion rate associated with the radial mass transport that is more or less consistent with 
Li data in solar twins. The more realistic value would be closer to $2\times 10^3$\,cm$^2$\,s$^{-1}$ (dot-dashed curve in Fig.~\ref{fig:f3}).
However, the observational data do not exclude the possibility that $\nu_{\rm v}$ could be as large as $3\times 10^4$\,cm$^2$\,s$^{-1}$ during the first
300 Myr and then reduced to $1.75\times 10^3$\,cm$^2$\,s$^{-1}$ (dotted curve in Fig.~\ref{fig:f3}; however, our rotation evolution
computations show that this variable viscosity does not solve the problem, as is illustrated by Fig.~\ref{fig:f8}). 
In other words, our assumed constant value for $\nu_{\rm v}$
should instead be considered a time average, therefore it can be a bit larger, say like 
$\nu_{\rm h}=4\times 10^3$\,cm$^2$\,s$^{-1}$ (short-dashed curve in Fig.~\ref{fig:f3}).
But it definitely cannot be as large as $\nu_{\rm v} = 5\times 10^4$\,cm$^2$\,s$^{-1}$ (solid curve in Fig.~\ref{fig:f3}) and remain constant, as CHMG93 assumed.
As for our choice of $\nu_{\rm h}$, its amplitude seems to have a correct order of magnitude (in agreement with other most recent applications)
for a ZAMS solar model rotating with the angular velocity $\Omega_{\rm e,ZAMS} = 100\,\Omega_\odot$, taking into account that \cite{mea07} have estimated
$\nu_{\rm h}\approx 10^7$\,--\,$10^8$\,cm$^2$\,s$^{-1}$ for a young (252 Myr) solar-metallicity $1.5\,M_\odot$ model star that started
its MS evolution with approximately one half the rotation rate. On the other hand, the assumed dependence of $\nu_{\rm h}$
on $\Omega_{\rm e,ZAMS}$ does not seem unnatural and it provides us with the necessary additional
degree of freedom, such that a variation of its value can probably be used to model the transition from the fastest to slowest rotators, without
causing a conflict with other observational constraints.

Thick solid curves in Fig.~\ref{fig:f6} have been computed using the anisotropic diffusion coefficients (\ref{eq:nuvnuh})
for the D2 poloidal field configuration. We have assumed that $\eta_{\rm v} = 2.5\times 10^3$ cm$^2$\,s$^{-1}$ and $\eta_{\rm h} = \nu_{\rm h}$.
The poloidal field decay has been taken into account (equation \ref{eq:potential}) with an appropriate value of the decay time
obtained as a solution of equation (\ref{eq:eigdim}). We can see that now the surface rotation evolution curve in panel A fits
the upper 90th percentiles of the $\Omega_{\rm e}$ distributions for solar analogs in open clusters, while its corresponding internal rotation profile
at the solar age appears to be nearly uniform (panel B). A comparison of four similar models in panel A
demonstrates that it is the combination of the D2 poloidal field geometry and strong horizontal turbulence (the D2H model) that results in
the rotational evolution resembling that of the fastest rotators because neither the D2 nor the D3H models are able to produce
a convex $\Omega_{\rm e}$ evolutionary curve. The contours of constant angular velocity and toroidal magnetic field for the D2H model
of the solar age are shown in panels B and D in Fig.~\ref{fig:f7}. It is important to note that,
in the absence of internal large-scale magnetic fields, strong horizontal turbulent diffusion could not erase the differential
rotation in the radiative core because it is the dipole poloidal field and its interaction with the rotational shear that
change the geometry of differential rotation from the spherically symmetric one (Fig.~\ref{fig:f7}C) to the toroidal one
(Fig.~\ref{fig:f5}D). On the other hand, the assumption of anisotropic turbulence with $\nu_{\rm h}\gg\nu_{\rm v}$ allows us
to choose a sufficiently small value of $\nu_{\rm v}$, thus avoiding a conflict with the observed surface Li abundances
in solar twins. 

\begin{figure}
\epsfxsize=14cm
\epsffile [60 180 480 695] {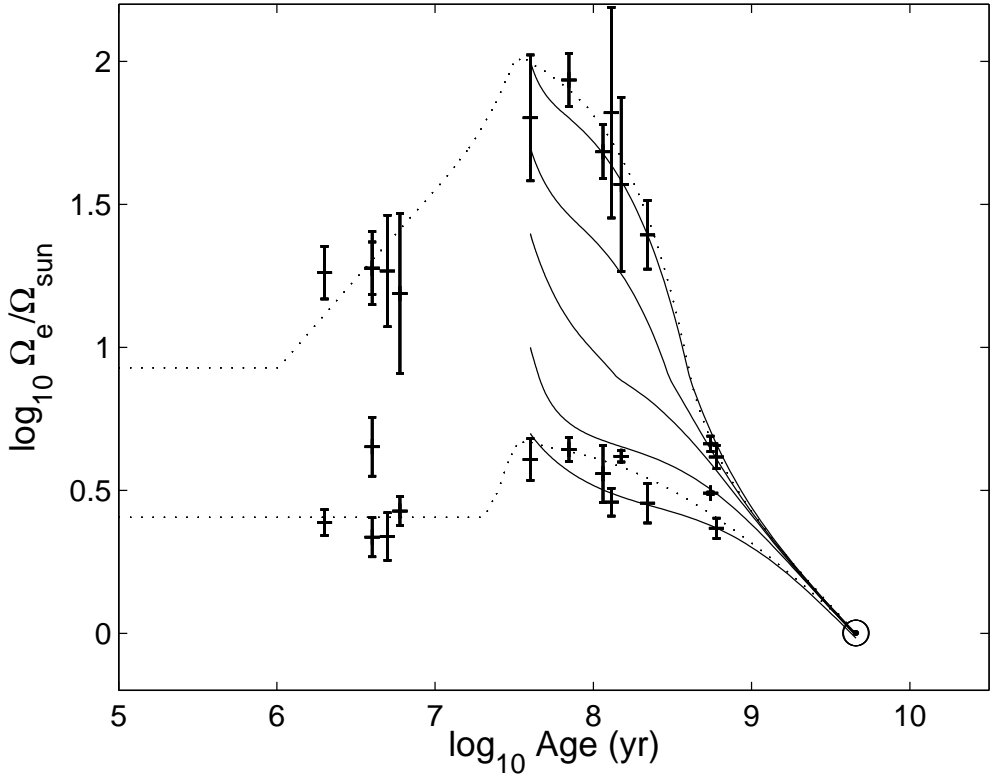}
\caption{Solid curves show the rotational evolution of solar-type stars computed with our D2H model using the anisotropic turbulent viscosity (\ref{eq:nuvnuh}).
         Other model parameters are $\eta_{\rm v} = 2.5\times 10^3$ cm$^2$\,s$^{-1}$ (except for the slowest rotating model
         that has $\eta_{\rm v} = 5\times 10^3$ cm$^2$\,s$^{-1}$), $\eta_{\rm h} = \nu_{\rm h}$, $B_0 = 0.1$ G initially and is decaying exponentially
         with an e-folding time that is calculated using equation (\ref{eq:eigdim}). The solar-calibrated wind constant
         varies between $K_{\rm w} = 3.64\times 10^{47}$ cm$^2$\,g\,s and $K_{\rm w} = 4.44\times 10^{47}$ cm$^2$\,g\,s from the fastest to slowest
         rotator.
         }
\label{fig:f9}
\end{figure}

Finally, when we apply the D2H configuration to model the rotational evolution of solar-type stars with
initial angular velocities ranging from $\Omega_{\rm e,ZAMS} = 100\,\Omega_\odot$ to $\Omega_{\rm e,ZAMS} = 5\,\Omega_\odot$,
we indeed obtain a nice transition from the convex evolutionary curves for the fastest rotators to the concave ones for the slowest
rotators (Fig.~\ref{fig:f9}), as hinted by observations. Note that all of these models end up with nearly flat internal
rotation profiles, and their solar-calibrated wind constants do not differ by very much: whereas the fastest rotating model has
$K_{\rm w} = 3.64\times 10^{47}$ cm$^2$\,g\,s, the slowest rotating model needs $K_{\rm w} = 4.44\times 10^{47}$ cm$^2$\,g\,s.

\subsection{Stationary Solutions: the Width of the Solar Tachocline}
\label{sec:tacho}

The conclusion made by \cite{mgch99} that it is the D3 configuration, rather than the D2 one, 
that seems to be the most appropriate one for the model of magnetic braking of solar rotation
is based on their consideration of the thinness of the solar tachocline. Because of the direct magnetic coupling
of the convective envelope with the radiative core through the D2 poloidal field, the latitudinal differential rotation at the core-envelope
interface $\Omega(r_{\rm e},\theta,t_\odot)$ penetrates into the bulk of the radiative core more easily in the D2 case than in the D3 case.
The differential rotation of the convective envelope in the present-day Sun is maintained by the inhomogeneous turbulence of a rotating fluid   
(e.g., \citealt{kr93,rk96}). Following \cite{chea99} and \cite{rk07}, we use the simple approximate relation
\bea
\Omega(r_{\rm e},\theta,t_\odot) = \Omega_\odot\,(1-0.15\cos^2\theta).
\label{eq:diffrot}
\eea
Its values at $\theta = 30^\circ$, $45^\circ$, $60^\circ$, and $90^\circ$ are plotted in panels C and D in Fig.~\ref{fig:f2} as
horizontal line segments at $r/R\ga 0.713$. It is seen that the variation of the interface angular velocity with the colatitude
in the present-day Sun is much smaller than the decrease of $\Omega_{\rm e}$ during the Sun's possible previous MS evolution (Fig.~\ref{fig:f2}A).
Therefore, we have neglected this variation in our rotational evolution computations. However, now, when we consider the final
model of the solar age, we have to be sure that our assumptions and, in particular, our preference for the D2 poloidal field configuration,
do not lead to the formation of a thick tachocline.

A stationary solution that has to be tested for the spread of the interfacial differential rotation can easily be obtained
with the same COMSOL Multiphysics code that we have used to compute the rotational evolution of solar-type stars. We have only to set
to zero the left-hand sides of equations (\ref{eq:mom})\,--\,(\ref{eq:ind}), omit the term proportional to $S_J$, use
the relation (\ref{eq:diffrot}) as a new outer boundary condition, and shrink the computational space domain from 
$r=R_\odot$ to $r=r_{\rm e}=0.713\,R_\odot$. We have used this approach to obtain a number of stationary solutions,
the results of which are discussed below.

\begin{figure}
\epsfxsize=12cm
\epsffile [10 120 480 695] {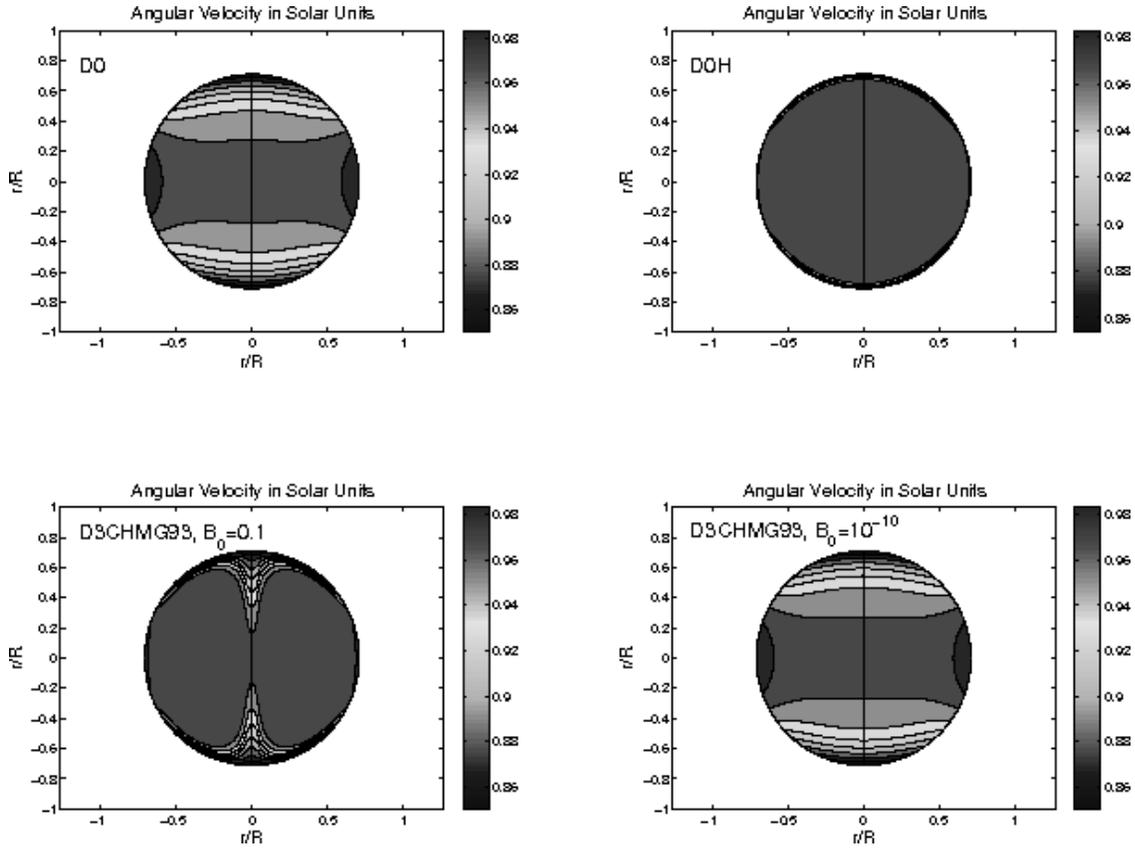}
\caption{Penetration of the latitudinal differential rotation of the bottom of Sun's convective envelope, 
         $\Omega(r_{\rm e},\theta,t_\odot)/\Omega_\odot = (1-0.15\cos^2\theta)$ (\citealt{rk07}),
         into its radiative core. D0: $\nu = 20$ cm$^2$\,s$^{-1}$ has its near
         maximum microscopic value. D0H: same as D0 but with the horizontal components of $\nu_{\rm h} = 10^4$ cm$^2$\,s$^{-1}$, note the formation of
         a thin tachocline in this case, as originally proposed by \cite{sz92}. 
         D3CHMG93, $B_0 = 0.1$ (G): the original D3 model of CHMG93 with the constant poloidal field (cf. lower right panel
         in Fig.~1 from \citealt{mgch99}). D3CHMG93, $B_0 = 10^{-10}$ (G): same as in the previous panel but with a strongly decayed poloidal field amplitude,
         which had not been taken into account by CHMG93.
         }
\label{fig:f10}
\end{figure}

Let us start with non-magnetic solar models. Fig.~\ref{fig:f10}A shows a stationary solution (contours of constant angular velocity)
for one such model in which the kinematic viscosity is isotropic and has its near maximum microscopic value of
$\nu = 20$ cm$^2$\,s$^{-1}$ that is assumed (as before) to be constant throughout the radiative core.
This solution is evidently in conflict with helioseismology data indicating that both the latitudinal and radial differential rotation
of the solar convective envelope are smoothed out when one crosses the thin ($\Delta r\la 0.04\,R_\odot$) tachocline, so that
the radiative core rotates like a solid body at least down to $r\approx 0.2\,R_\odot$ (e.g., \citealt{rk07}, and references therein).
This result is not surprising because the isotropic viscosity is known to spread a horizontal velocity inhomogeneity of a given length scale
to a radial velocity inhomogeneity of the same length scale (\citealt{rk97}).
Fig.~\ref{fig:f10}B presents a stationary solution for the same non-magnetic model but with the anisotropic viscosity whose horizontal component is
$\nu_{\rm h} = 10^4$ cm$^2$\,s$^{-1} \gg \nu_{\rm v} = 20$ cm$^2$\,s$^{-1}$. In this case, the strong horizontal turbulence
smooths out the interfacial differential rotation in a sufficiently thin layer located immediately beneath the core-envelope interface
(cf. \citealt{sz92}).

Now, we return to the isotropic viscosity but add a poloidal magnetic field. Fig.~\ref{fig:f10}C
shows a stationary solution for the case similar to that considered by CHMG93, namely for the D3 poloidal field configuration
of amplitude $B_0 = 0.1$ G interacting with the rotational shear via the (turbulent) viscosity $\nu = 5\times 10^4$ cm$^2$\,s$^{-1}$,
the induced toroidal field being diffused away at the rate $\eta = 5\times 10^5$ cm$^2$\,s$^{-1}$.
In this case, our contours of constant angular velocity indeed resemble very closely those plotted in the lower right panel in Fig.~1 by \cite{mgch99}.
After returning to the isotropic viscosity, we would expect the interfacial differential rotation to again be spread radially
deep into the core but this has not happened. The reason is that now it is the latitudinal transport of angular momentum by the magnetic stress,
proportional to the product $B_\theta B_\varphi$, that suppresses the differential rotation and restricts its penetration into
the radiative interior (\citealt{rk97}). However, the original CHMG93 model did not take into account the decay of the poloidal field in spite of the fact that
its assumed extremely high value of the magnetic diffusivity implies the very short decay time of 35 Myr. With this e-folding time,
the poloidal field will already have the amplitude of merely $10^{-10}$ G by the age of 725 Myr. Fig.~\ref{fig:f10}D
shows contours of constant angular velocity in the D3 model of the solar age with this weak field. Because there is actually no poloidal field
left, the solution looks almost identical to that for the pure D0 case.

\begin{figure}
\epsfxsize=12cm
\epsffile [10 120 480 695] {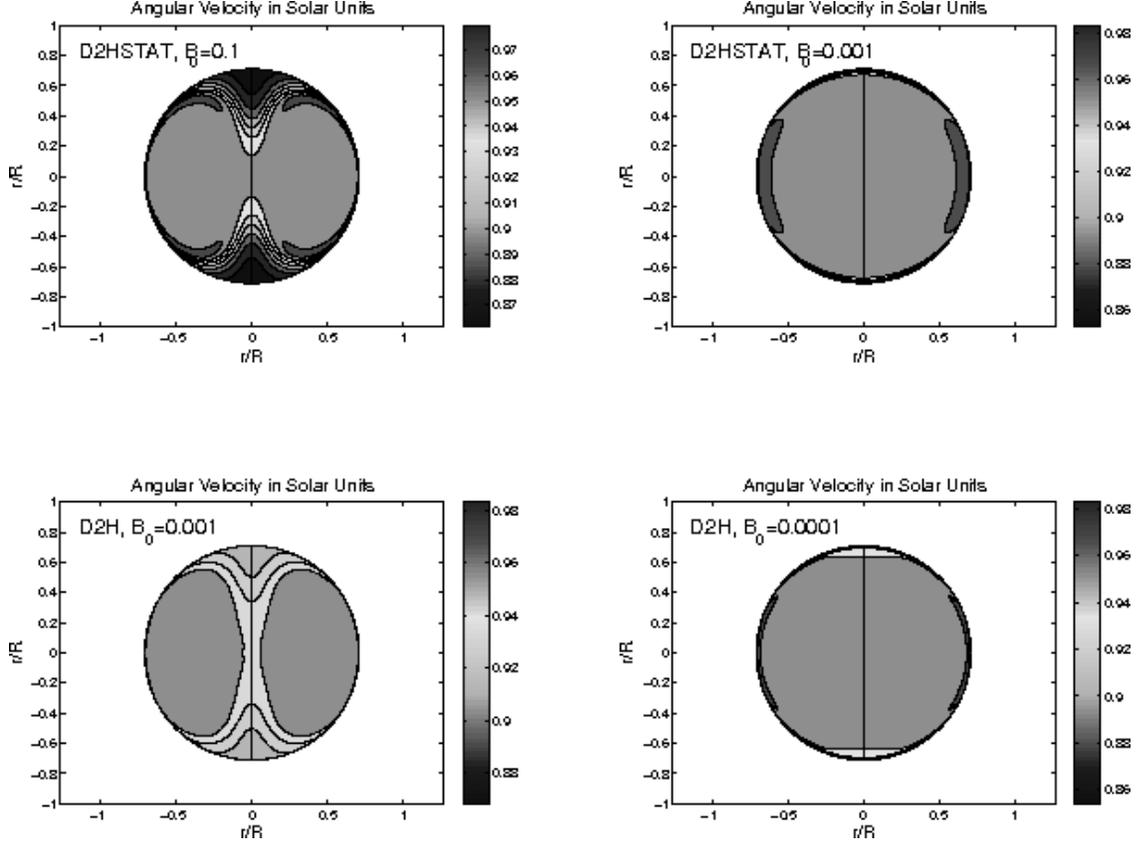}
\caption{Same as in Fig.~\ref{fig:f10} but for our magnetic braking model (the D2 poloidal field with the anisotropic turbulent diffusion).
         D2HSTAT, $B_0 = 0.1$ (G): this case assumes that all the model parameters remain constant, including the value of $\nu_{\rm h} = 10^6$ cm$^2$\,s$^{-1}$
         (for $\Omega_{\rm e,ZAMS} = 100\,\Omega_\odot$ in equation \ref{eq:nuvnuh}). D2HSTAT, $B_0 = 0.001$ (G): same as in the previous panel but
         with the decayed poloidal field. D2H, $B_0 = 0.001$ (G) and $B_0 = 0.0001$ (G): here, it is additionally assumed that $\nu_{\rm h}$ has been reduced
         to $10^4$ cm$^2$\,s$^{-1}$, proportionally to the decrease of $\Omega_{\rm e}$ from $100\,\Omega_\odot$ to $\Omega_\odot$.
         }
\label{fig:f11}
\end{figure}

We have chosen the D2 configuration because it more effectively transfers the wind torque and rotational shear into the radiative core than
the D3 configuration. However, this same property plays a negative role when we try to prevent the interfacial differential rotation
in the present-day Sun from penetrating the radiative interior. Fortunately, our D2H model combines the D2 geometry with the assumption of
strong horizontal turbulent diffusion which, as we have seen in Fig.~\ref{fig:f10}B, can potentially counteract the negative effect produced
by the non-vanishing radial component of the D2 poloidal field at the core-envelope interface. 
To test this possibility, we have computed four stationary solutions
for our favorite D2H model. Fig.~\ref{fig:f11}A shows contours of constant angular velocity at the solar age for a D2H model in which
neither the anisotropic diffusion coefficients nor the poloidal field amplitude were allowed to change with time. This is the worst possible solution
in terms of the tachocline width. However, like the original CHMG93 solution, our first D2H test solution is not fully consistent because it does not take into account
the poloidal field decay. For our assumed values of $\eta_{\rm v} = 2.5\times 10^3$ cm$^2$\,s$^{-1}$ and
$\eta_{\rm h} = \nu_{\rm h} = 10^6$ cm$^2$\,s$^{-1}$ (for the fastest solar-type rotator with $\Omega_{\rm e,ZAMS} = 100\,\Omega_\odot$)
equation (\ref{eq:eigdim}) gives $\tau = 1.1$ Gyr. Hence, by the solar age, the initial amplitude $B_0 = 0.1$ G
will be reduced to $2\times 10^{-3}$ G. Fig.~\ref{fig:f11}B demonstrates that for a D2 poloidal field of such an order of
magnitude the tachocline can be very thin, due to the continuing strong horizontal turbulent diffusion. 
Now, given our assumption about the dependence of $\nu_{\rm h}$ on the angular velocity (equation \ref{eq:nuvnuh}),
it would be fair to construct a solution for the case of $\nu_{\rm h}$ having been reduced to $10^4$ cm$^2$\,s$^{-1}$, proportionally to
the decrease of $\Omega_{\rm e}$ from $100\,\Omega_\odot$ to $\Omega_\odot$. Its respective contours of constant $\Omega(r,\theta,t_\odot)$
are presented in Fig.~\ref{fig:f11}C. We see that even in this case the obtained tachocline width is not more discrepant than in the original CHMG93 solution
(Fig.~\ref{fig:f10}C). Finally, Fig.~\ref{fig:f11}D illustrates that the previous solution is greatly improved
when the amplitude of the residual poloidal field is reduced by another order of magnitude. So, the anisotropic turbulence
with the horizontal components of turbulent viscosity strongly dominating over those in the vertical direction plus
the account of the poloidal field decay incorporated into our D2H model make it possible to obtain a sufficiently thin tachocline
in the present-day Sun's model even for the D2 poloidal field configuration.

\section{Conclusion}
\label{sec:concl}

The main goal of this work has been to modify the model of magnetic braking of solar rotation presented and discussed in detail
by CHMG93 so that it could reproduce the most recent observational data concerning the rotation of solar analogs in open clusters of different ages,
without coming into conflict with other available observational constraints. Although we have used a different prescription for
the surface angular momentum loss rate (equation \ref{eq:jdot}), its corresponding dependence of the envelope braking rate
$-d(\Omega_{\rm e}/\Omega_\odot)/dt$ (yr$^{-1}$) on the rotation rate $\Omega_{\rm e}/\Omega_\odot$ does not deviate significantly from
the dependence predicted by the Weber-Davis MHD wind model that was employed by CHMG93 (for comparison, we used
Fig.~1 from \citealt{ch92}). It turns out that even in its original formulation the CHMG93 model experiences difficulties
in fitting the upper 90th percentiles of the $\Omega_{\rm e}$ distributions for cluster solar-type stars. The discrepancy
with the observationally constrained rotational evolution of the fastest rotators becomes much worse when we reduce
the kinematic viscosity by one order of magnitude and/or take into account the poloidal field decay caused by the high
magnetic diffusivity in the CHMG93 model. The viscosity reduction makes the model more consistent with the data for
the atmospheric Li depletion in solar twins. The value of $\nu = 5\times 10^4$ cm$^2$\,s$^{-1}$ used by CHMG93
leads to excessive Li destruction unless it describes, in terms of an effective diffusion coefficient,
another angular momentum transport mechanism implemented as a process without a radial mass transfer, 
e.g. angular momentum redistribution by  internal gravity waves.
However, in this case the question would arise as to why that additional process could not do the entire work alone?
Poloidal field decay also seems to be necessary, given that the model already follows the decay of the induced
toroidal field. After these reasonable modifications are made, the CHMG93 model produces a rotational evolution of
a rapidly rotating solar-type star that differs greatly from that of the fastest rotators among solar analogs
in open clusters. Besides, at the solar age, the model still possesses strong differential rotation in the radiative core, in conflict
with helioseismology measurements. This differential rotation has such a geometry that its corresponding contours of constant
angular velocity form consecutive layers in cross sections of an axisymmetric torus, a region called the ``dead zone''
by \cite{mw87}. Its shape is a manifestation of Ferraro's law of isorotation, $(\mathbf{B}_{\rm p},\nabla)\Omega = 0$,
with the poloidal field $\mathbf{B}_{\rm p}$ having a dipole structure.

To save the CHMG93 model, we have assumed that the rotation-driven turbulence that is thought to amplify the viscosity and
magnetic diffusivity compared to their microscopic values in stellar radiative zones is actually anisotropic with the horizontal
components of the transport coefficients strongly dominating over those in the vertical direction. This is not a new hypothesis.
It was first introduced to the stellar astrophysics community by \cite{z92} and, since then, it has widely been used to
model both chemical mixing and angular momentum transport in stars. The strong horizontal turbulent diffusion 
serves a two-fold goal in our modified model: it erases the dead zone along isobaric surfaces (a horizontal erosion of latitudinal differential rotation),
while allowing us to choose a sufficiently small value for the coefficient of vertical diffusion that does not lead to a conflict
with the Li data in solar twins. We have also found it necessary to switch from the D3 poloidal field configuration,
that was considered to be the most appropriate one in the original CHMG93 model, to the D2 configuration.
In the D2 geometry, the dipole poloidal field partially penetrates the convective envelope, which allows it to more effectively
transfer the surface wind torque to the radiative interior. As a result, a thinner dead zone develops, which makes it
easier for the horizontal turbulence to erase it. The horizontal component of the turbulent viscosity gives us
an additional degree of freedom whose assumed linear dependence on the initial (ZAMS) angular velocity
enables the models to properly reproduce the transition from the convex morphology of the surface rotation evolution of the fastest rotators
to the concave $\Omega_{\rm e}(t)$ curves for the slowest rotators, as suggested by observations.

\cite{mgch99} have chosen the D3 geometry as the most suitable one for the modeling of magnetic breaking of solar rotation
because its corresponding poloidal field configuration
does not lead to a penetration of the differential rotation of the Sun's convective envelope into its radiative interior, consistent with
the observations. The helioseismology measurements show that the envelope differential rotation gets smoothed out in a thin ($\Delta r\la 0.04\,R_\odot$)
transition layer (the solar tachocline) immediately beneath the bottom of the Sun's convective envelope and that the radiative core
below the tachocline rotates like a solid body at least down to $r\approx 0.2\,R_\odot$. Contrary to this, it has been shown that
the non-vanishing radial component of the D2 poloidal field at the core-envelope interface causes
the interfacial latitudinal differential rotation to move into the bulk of the radiative core. However, after our modifications of
the original CHMG93 model, the new D2H model appears to be free of this problem. Both the strong horizontal turbulent diffusion and the poloidal field decay
work towards confining the width of the solar tachocline.

It is important to note that our D2H model is very simplistic and it is based on a number of assumptions whose validity has not been proven.
For instance, it is not clear how the rotation-driven turbulence should interact with the strong large-scale magnetic fields.
Also, the formation of the solar tachocline is a much more complex process than described by our model. In particular,
it must take into account a penetration of the meridional circulation from the convective zone into the radiative
core (e.g., \citealt{gmi98,rk07,g09}), which our model completely ignores.
We are aware of these shortcomings. Therefore, we consider our model only as a legitimate extension of the CHMG93 model
in the sense that we have not made any modifications of the latter that are very different from the original assumptions and approximations used by CHMG93.
On the other hand, we have shown that just one additional assumption, that of strong horizontal viscosity,
helps to solve several problems with the old model.

\acknowledgements
The author is grateful to Don VandenBerg who has supported this work through his Discovery Grant
from Natural Sciences and Engineering Research Council of Canada. The author also appreciates
discussions with Marc Pinsonneault, Don Terndrup, and Keith MacGregor that have greatly stimulated this work.



\clearpage

\end{document}